\documentclass[journal]{IEEEtran}
\usepackage{amssymb}
\usepackage{stfloats}
\usepackage{graphicx}
\usepackage{multirow}
\usepackage{subfigure}
\usepackage{tabularx}
\usepackage{booktabs}
\usepackage{amsmath}
\usepackage{amsthm}
\usepackage{epsfig,epsf,color,balance,cite}
\usepackage{indentfirst}
\usepackage{mathrsfs}
\usepackage{dsfont}
\UseRawInputEncoding
\usepackage{easyReview}
\usepackage[justification=centering]{caption}
\usepackage[font=small,labelfont=md,labelsep=period]{caption}

\newtheorem{remark}{Remark}
\usepackage{algorithm, algorithmicx, algpseudocode}
\usepackage{enumerate}
\usepackage{makecell}
\newcounter{MYtempeqncnt}
\usepackage[colorlinks,bookmarksopen,bookmarksnumbered,citecolor=red, linkcolor=blue, urlcolor=blue]{hyperref}

\makeatletter
\DeclareRobustCommand{\cev}[1]{%
	{\mathpalette\do@cev{#1}}%
}
\newcommand{\do@cev}[2]{%
	\vbox{\offinterlineskip
		\sbox\z@{$\m@th#1 x$}%
		\ialign{##\cr
			\hidewidth\reflectbox{$\m@th#1\vec{}\mkern4mu$}\hidewidth\cr
			\noalign{\kern-\ht\z@}
			$\m@th#1#2$\cr
		}%
	}%
}
\makeatother

\begin{document}
	\title{Spatially Non-Stationary XL-MIMO Channel Estimation: A Three-Layer Generalized Approximate Message Passing Method}
	\author{
		\IEEEauthorblockN{ 
			Anzheng Tang\IEEEauthorrefmark{1}, ~\IEEEmembership{Graduate Student Member,~IEEE},
			Jun-Bo Wang\IEEEauthorrefmark{1}, ~\IEEEmembership{Senior Member,~IEEE},\\
			Yijin Pan\IEEEauthorrefmark{1}, ~\IEEEmembership{Member,~IEEE},
			Wence Zhang\IEEEauthorrefmark{2}, ~\IEEEmembership{Member,~IEEE},
			Yijian Chen\IEEEauthorrefmark{3},
			Hongkang Yu\IEEEauthorrefmark{3},\\
			and Rodrigo C. de Lamare\IEEEauthorrefmark{4}}, ~\IEEEmembership{Senior Member,~IEEE}
		\thanks{This work was supported in part by the National Natural Science Foundation of China under Grant No. 62371123, 6504009793 and 6504009805, in part by ZTE Industry-University-Institute Cooperation Funds under Grant No. IA20231212016, and in part by Research Fund of National Mobile Communications Research Laboratory Southeast University under Grant No. 2024A03. (\emph{Corresponding authors: Jun-bo Wang and Yijin Pan})}
		\thanks{A. Tang, J.-B. Wang and Y. Pan are with the National Mobile Communications Research Laboratory, Southeast University, Nanjing 210096, China. (E-mail: \{anzhengt, jbwang, and panyj\}@seu.edu.cn)}
		\thanks{W. Zhang is with the School of Computer Science and Communication Engineering, Jiangsu University, Zhenjiang 212013, China. (E-mail: wencezhang@ujs.edu.cn)}
		\thanks{Y. Chen and H. Yu are with the Wireless Product Research and Development Institute, ZTE Corporation, Shenzhen 518057, China. (E-mail:\{yu.hongkang, chen.yijian\}@zte.com.cn)}
		\thanks{Rodrigo C. de Lamare is with the Centre for Telecommunications Studies, Pontifical Catholic University of Rio de Janeiro, Rio de Janeiro 22451-900, Brazil, and also with the School of Physics, Engineering and Technology, University of York, York YO10 5DD, U.K. (E-mail: delamare@puc-rio.br).}
	}
	\maketitle
	\begin{abstract}
		In this paper, the channel estimation problem for extremely large-scale multi-input  multi-output (XL-MIMO) systems is investigated with the considerations of near-field (NF) spherical wavefront effects and spatially non-stationary (SnS) properties.
		Due to the diversity of SnS characteristics across different propagation paths, the concurrent channel estimation of multiple paths becomes intractable. 
		To address this challenge, we propose a two-phase estimation scheme that decouples the problem into multiple subchannel estimation tasks. 
		To solve these sub-tasks, we introduce a novel three-layer Bayesian inference scheme, exploiting the correlations and sparsity of the SnS subchannels in both the spatial and angular domains. 
		Specifically, the first layer captures block sparsity in the angular domain, the second layer promotes SnS properties in the spatial domain, and the third layer effectively decouples each subchannel from the observed signal. To enable efficient Bayesian inference, we develop a three-layer generalized approximate message passing (TL-GAMP) algorithm that combines structured variational message passing with belief propagation rules.
		Simulation results validate the convergence and effectiveness of the proposed TL-GAMP algorithm, demonstrating its robustness across various channel environments, including NF-SnS, NF spatially stationary (NF-SS), and far-field spatially stationary (FF-SS) scenarios.
	\end{abstract}   
	\IEEEpeerreviewmaketitle
	\section{Introduction}
	\label{section1}
	With the advancement of wireless communication technology and the growing demand for higher data rates, it is anticipated that the number of antennas and the array aperture will significantly exceed those used in existing massive multiple-input-multiple-output (MIMO) systems \cite{6G1}. This trend has given rise to the concept of extremely large-scale MIMO (XL-MIMO), which involves deploying an exceptionally large number of antennas in a compact space with a discrete or even continuous aperture. Thanks to the substantial beamforming gain and vast spatial degrees of freedom (DoFs), XL-MIMO is considered a promising technology for beyond 5G and 6G communications\cite{XL_MIMO3, XL_MIMO_T1}.
	
	The electromagnetic (EM) radiation field emitted by antennas is traditionally categorized into far-field (FF) and near-field (NF) regions \cite{XL_MIMO_T2}. In conventional massive MIMO systems, due to the limitations of array aperture and operating frequency bands, users are typically located in the FF region. However, in XL-MIMO systems, the deployment of extremely large antenna arrays (ELAAs) and the use of higher frequency bands enable NF communications to be effective over distances of hundreds of meters \cite{LoS_Angular}. In this manner, the FF plane wavefront assumption becomes invalid for NF XL-MIMO channels, necessitating the consideration of spherical wavefront effects \cite{XL_MIMO5}. 
	Furthermore, the significantly larger array aperture introduces spatially non-stationary (SnS) properties \cite{SnS1, SnS2, Non_Stationary_Fading}. 
	Unlike spatially stationary (SS) massive MIMO channels, XL-MIMO channels allow different portions of the array to observe the propagation environment from different perspectives, meaning that array elements can receive signals from the same propagation path but with varying power levels. Consequently, considering both NF effects and SnS properties, channel estimation for XL-MIMO systems becomes significantly more challenging.
	\vspace{-1em}
	\subsection{Related Works}
	Due to the randomness of user positions and the dynamic scattering environment, XL-MIMO channels can be classified into three primary scenarios: NF-SnS, NF-SS, and FF-SS.
	\begin{itemize}
		\item {NF-SnS: Users or scatterers are located in the NF region, with only a subset of antennas in the antenna array able to observe the users or scatterers for each propagation path. As a result, each user's power is concentrated within a specific region of the array.}
		\item {NF-SS: Users or scatterers are also situated in the NF region. However, both users and scatterers are visible to all antennas in the array for each propagation path.}
		\item {FF-SS: Users and scatterers are positioned in the FF region, where the distance between the users and the array is significantly greater than the aperture of the array. Thus, users are fully observed by the array, which is consistent with the conventional massive MIMO channels.}
	\end{itemize}
	
	{While several methods have been proposed for channel estimation in XL-MIMO systems, most of these focus on NF-SS scenarios to address the spherical wavefront effects} \cite{PolarCS, NFBT1, NFBT2, NFCE1, NFCE2}. Recently, to address the NF-SnS channel estimation problem, \cite{Hanyu2} proposed subarray-wise and scatterer-wise estimation schemes. However, these schemes were heuristic and did not fully exploit the inherent structure of the NF-SnS channels, such as potential characteristics in the spatial or angular domains. 
	Moreover, based on the assumption of subarray-wise visibility regions (VRs), \cite{VR_wise_SnS} proposed a group time block code (GTBC) based signal extraction scheme for each subarray. {Subsequently, the NF-SnS channel estimation was transformed into several FF-SS estimation tasks.} 
	Nevertheless, the method overlooked the spatial correlation among subarrays, and the subarray-wise VR assumption may be idealistic, as users or scatterers might only have visibility to a portion of antennas in each subarray.
	Additionally, in the context of SnS reconfigurable intelligent surface (RIS) cascaded channels, a three-step channel estimation and VR detection scheme was proposed in \cite{HanYu}, where VR detection method relies on received signal power, is contingent on channel estimation accuracy, and is sensitive to noise levels. Particularly, in low signal-to-noise ratio (SNR) scenarios, this approach may suffer severe performance degradation. Furthermore, the method becomes impractical when considering multipath propagation between the RIS and users.
	
	{Considering the inherent spatial-domain or delay-domain sparsity in NF-SnS channels, various Bayesian inference-based methods have been proposed.} Exploiting delay-domain sparsity, \cite{SnS4} introduced an adaptive grouping sparse Bayesian learning (AGSBL) scheme for uplink channel estimation. Utilizing spatial-domain sparsity resulting from the SnS properties, \cite{Bayesian3} characterized XL-MIMO channels with a subarray-wise Bernoulli-Gaussian distribution. Subsequently, a bilinear message passing (MP) algorithm was developed for joint user activity detection and channel estimation in the presence of the SnS properties. To simultaneously capture spatial- and delay-domain sparsity, \cite{Bayesian1} proposed a structured prior model with the hidden Markov model (HMM) to capture the characteristics of   VR and delay domain clustering. While these contributions primarily focused on characterizing spatial-domain or delay-domain channels using statistical distributions such as mean and variance, they overlooked essential propagation characteristics, such as the angular- or wavenumber-domain properties, which potentially lead to a degradation in estimation performance.
	
	{To address the aforementioned issues, \cite{RIS_VR} proposed to jointly utilize the sparse properties of NF-SnS channels in the polar and spatial domains, developing a joint channel estimation and VR detection algorithm based on the fast sparse Bayesian learning (FSBL) framework.}
	Specifically, tailored to the SnS properties, \cite{RIS_VR} proposed an add-on model, where the visibility indicator vector is represented as a linear combination of a series of candidate blockage vectors, and the estimation of the VR vector is transformed into the estimation of combination coefficients. However, the add-on based VR representation is heavily dependent on prior information about the size of VRs, which limits its applicability. Particularly, when there is no prior information, the number of combination coefficients is much greater than the number of antennas. For XL-MIMO systems, this poses a significant challenge for computational costs.
	In addition, our previous work \cite{Turbo-OMP} proposed a two-stage VR detection and channel estimation scheme based on approximated MP (AMP) and orthogonal matching pursuit (OMP), considering the spatial- and angular-domain characteristics. Nevertheless, a notable drawback of this method is the manual separation of the two stages, leading to inadequate utilization of angular-domain information during the VR detection stage.
	\vspace{-1em}
	\subsection{Main Contributions}
	{To address the aforementioned challenges, this paper investigates the NF-SnS channel estimation problem for XL-MIMO systems, specifically considering a user equipped with multiple antennas.} The main contributions are summarized as follows:
	\begin{itemize}
		\item {Tailored to the diverse SnS characteristics across different propagation paths, we propose a novel two-phase channel estimation scheme. In the first phase, the angles of departure (AoDs) on the user side are estimated using a super-resolution method. In the second phase, leveraging the estimated AoDs, the overall channel estimation problem is decoupled into multiple subchannel estimation tasks.}
		\item {To accurately solve the cascaded estimation of VRs and channel coefficients in the second phase, we formulate the subchannel estimation as a Bayesian inference problem. Hierarchical-sparse and Markov-chain-based prior models are employed to characterize the SnS subchannel in the angular and spatial domains, respectively. However, the introduction of the Markov-chain structure presents significant challenges in the forward and backward message updates within existing parametric bilinear inference schemes \cite{Para_Bi_Est}.
		To address this issue, we propose a novel layered Bayesian inference scheme, where the angular channels and VR indicator vectors are estimated in separate layers, effectively avoiding the complex graph topology between variable and factor nodes.}
		\item {To efficiently achieve Bayesian inference, we propose a three-layer generalized approximate message passing (TL-GAMP) algorithm tailored for structured prior models. In this scheme, the first layer captures the block sparsity of the SnS channel within the angular domain. The second layer focuses on modeling the SnS properties in the antenna domain, while the third layer facilitates the decoupling of subchannels from the observed signal. Additionally, to reduce computational complexity, we simplify the message calculations in the first and third layers by omitting certain infinitesimal terms.}
		\item {Compared to existing approaches, the TL-GAMP algorithm achieves superior performance by jointly leveraging angular- and spatial-domain characteristics of SnS channels, enabling it to closely approach the lower bound established by perfect visibility knowledge. Additionally, the TL-GAMP algorithm exhibits exceptional robustness across diverse channel conditions, including NF-SnS, NF-SS, and FF-SS scenarios, underscoring its adaptability in various environments.}
	\end{itemize} 
	\vspace{-1em}
	\subsection{Organization and Notations}
	Organization: The rest of this paper is organized as follows. In Section \ref{section2}, we introduce the system model. Section \ref{section3} presents the two-phase channel estimation scheme. In Section~\ref{section4}, we formulate the subchannel estimation problem as a three-layer Bayesian inference problem and propose a computationally efficient algorithm. Simulations are presented in Section \ref{section5}. Finally, conclusions are drawn in Section \ref{section6}.
	
	Notations: lower-case letters, bold-face lower-case letters, and bold-face upper-case letters are used for scalars, vectors and matrices, respectively; 
	The superscripts $\left(\cdot \right)^{\mathrm{T}}$ and $\left(\cdot \right)^{\mathrm{H}}$ stand for transpose and conjugate transpose, respectively;  
	$\mathrm{diag}\left(x_1, x_2, \dotso, x_N\right)$ denotes a diagonal matrix with $\{x_1, x_2, \dotso, x_N\}$ being its diagonal elements; 
	$\mathrm{blkdiag}\left(\mathbf{X}_1, \mathbf{X}_2, \dotso, \mathbf{X}_N\right)$ denotes a block diagonal matrix with $\{\mathbf{X}_1, \mathbf{X}_2, \dotso, \mathbf{X}_N\}$ being its diagonal elements;  
	$\mathbb{C}^{M \times N}$ denotes an $M \times N$ complex matrix. 
	In addition, a random variable $x \in \mathbb{C}$ drawn from the complex Gaussian distribution with mean $m$ and variance $v$ is characterized by the probability density function (PDF) $\mathcal{CN}(x;m,v) = \exp(-{\left|x-m\right|^2}/{v})/{\pi v} $; {a random variable $\gamma \in \mathbb{R}$ from Gamma distribution with mean $a/b$ and variance $a/b^2$ is characterized by the PDF $\mathcal{G}a(\gamma; a, b) \propto \gamma^{a-1}\exp(-\gamma b)$;} $m_{n_i \rightarrow n_j}(x)$ indicates a message passed from node $n_i$ to node $n_j$.
	
	\begin{figure*}
		\centering
		\includegraphics[width=0.7\textwidth]{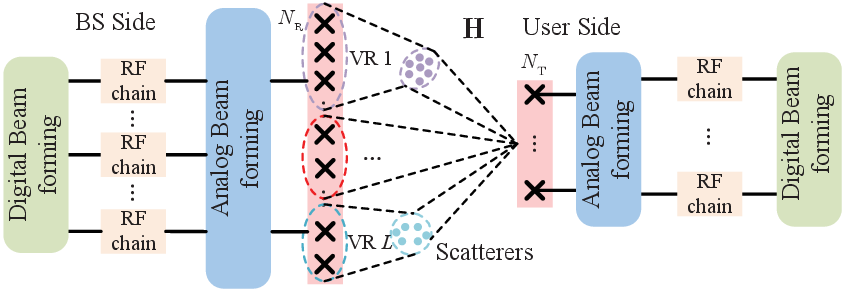}
		\caption{Illustration of a mmWave XL-MIMO system with a hybrid beamforming architecture.}
		\label{System Model}
	\end{figure*}
	
	\vspace{-1em}
	\section{System Model}
	\label{section2}
	{Consider a millimeter-wave (mmWave) XL-MIMO system where a base station (BS) with $N_{\mathrm{R}}$ antennas and $N_{\mathrm{RF}}$ radio frequency (RF) chains serves $J$ users in time-division duplexing (TDD) mode. 
	For uplink channel estimation, assume $J$ users transmit pilot sequences to the BS utilizing mutual orthogonal time sequences. Therefore, channel estimation for each user is independent each other. 
	Without loss of generality, we focus on the channel estimation problem for an arbitrary user.}
	As depicted in Fig.~\ref{System Model}, assume that each user is equipped with $N_{\mathrm{T}} < N_{\mathrm{R}}$ antennas, and the transceivers are arranged with uniform linear arrays (ULAs), where the antenna spacing is denoted by $d = \lambda/2$, with $\lambda$ indicating the carrier wavelength.
	In addition, the set of antennas for BS is given by {$\mathbb{N}_{\mathrm{R}} = \{1,2,\cdots, N_{\mathrm{R}}\}$}.
	
	{This paper aims to address the general problem of NF-SnS channel estimation for XL-MIMO systems. The proposed estimation algorithm is versatile and can be directly adapted for NF-SS and FF-SS scenarios, broadening its applicability across various XL-MIMO channel conditions.
	Considering the spherical wavefront effects and the SnS properties,} {a single-side NF-SnS model\footnote{{Due to size and power constraints, user equipment generally has far fewer antennas compared to the base station. Consequently, it is common and practical to model the base station steering vector under NF conditions and the user-side steering vector under FF conditions.}} is employed to characterize the uplink channel $\mathbf{H} \in \mathbb{C}^{N_{\mathrm{R}}\times N_{\mathrm{T}}}$. Mathematically, it can be represented as} \cite{SnS1, Bayesian1}
	\begin{equation}
		\begin{aligned}
			\mathbf{H} = \sqrt{N_\mathrm{T}N_\mathrm{R}}\sum_{l=1}^{L}g_{l}\left(\mathbf{s}_l \odot \mathbf{a}_{\mathrm{R}}(\vartheta_l, r_l)\right)\mathbf{a}_{\mathrm{T}}^{\mathrm{H}}(\psi_l),
		\end{aligned}
		\label{Channel}
	\end{equation}
	{where $L$ is the number of resolvable paths, and when $l = 1$, it refers to the line-of-sight (LoS) path, while $l > 1$ indicates non-LoS (NLoS) paths; $g_l$ represents the complex channel gain corresponding to the $l$-th path;} 
	{$\mathbf{s}_l = [s_{l,1}, s_{l,2}, \cdots, s_{l,N_\mathrm{R}}]^{\mathrm{T}}\in \{0, 1\}^{N_\mathrm{R}\times 1}$ is the visibility indicator vector of the $l$-th path\footnote{{
	Given the limited scattering, diffraction capabilities, and poor penetration of mmWave frequency bands, this paper adopts the visibility region (VR)-based channel model to simplify the characterization of SnS properties arising from different physical propagation mechanisms, following prior works \cite{HanYu,VR_wise_SnS,Turbo-OMP,SnS1}. In future research, we plan to explore more effective SnS channel estimation methods for broader and more general models, such as the one proposed in \cite{Non_Stationary_Fading}.}}; $\odot$ indicates the Hadamard product. If the $l$-th propagation path is a SS path, $\mathbf{s}_l$ is represented as an all-one vector, i.e., $\mathbf{s}_l = \mathbf{1}_{N_{\mathrm{R}} \times 1}$. 
	Conversely, if the $l$-th path is a SnS path, the $n$-th entry of $\mathbf{s}_l$ is modeled as~\cite{SnS1, Turbo-OMP}
	\begin{equation}
		s_{l,n} =
		\begin{cases}
			1, & \text{if } n \in \boldsymbol{\varphi}_l, \\
			0, & \text{otherwise}.
		\end{cases}
	\end{equation}
	{where set $\boldsymbol{\varphi}_l \subseteq  \mathbb{N}_{\mathrm{R}}$ denotes the VR for the $l$-th path, i.e., the $l$-th path is visible to elements in $\boldsymbol{\varphi}_l$, and is invisible to the elements outside $\boldsymbol{\varphi}_l$.} Moreover, we define $\phi_l = {\left|\boldsymbol{\varphi}_l\right|}/{N_{\mathrm{R}}}$, indicating the proportion of visible elements to the $l$-th path, where $\left|\boldsymbol{\varphi}_l\right|$ denotes the cardinality of the set $\boldsymbol{\varphi}_l$. The parameters $\vartheta_l$ and $r_l$ denote the angle of arrival (AoA) and the distance between the reference antenna element on the BS side and the scatterers or user, respectively. Without loss of generality, we consider the first antenna element as the reference element. Additionally, $\psi_l$ represents AoD of the user in the $l$-th path. Finally, $\mathbf{a}_{\mathrm{T}}(\psi_l) \in \mathbb{C}^{N_{\mathrm{T}} \times 1}$ and $\mathbf{a}_{\mathrm{R}}(\vartheta_l, r_l) \in \mathbb{C}^{N_{\mathrm{R}} \times 1}$ are the transmit and receive array steering vectors, which are respectively given by
	\begin{align}
		\label{A_t} \mathbf{a}_{\mathrm{T}}(\psi_l) &= \frac{1}{\sqrt{N_{\mathrm{T}}}}[1, \cdots, \mathrm{e}^{-\mathrm{j}\frac{2\pi}{\lambda}(N_{\mathrm{T}}-1)d\sin\psi_l}]^{\mathrm{T}}, \\
		\label{A_r} \mathbf{a}_{\mathrm{R}}(\vartheta_l, r_l) &= \frac{1}{\sqrt{N_{\mathrm{R}}}}[\mathrm{e}^{\mathrm{j}\frac{2\pi}{\lambda}\Delta_{l,1}}, \cdots, \mathrm{e}^{\mathrm{j}\frac{2\pi}{\lambda}\Delta_{l,N_{\mathrm{R}}}}]^{\mathrm{T}},
	\end{align}
	where $\Delta_{l,n}$ is the wave path difference between the $n$-th antenna and the reference antenna for the $l$-th path, which is given by $\Delta_{l,n}{\approx} -d(n-1)\sin \vartheta_l + {d^2(n-1)^2 \cos^2\vartheta_l}/{2r_l}$
	where $r_{l,n}$ denotes the distance between the $l$-th scatterer and the $n$-th receive antenna, where the approximation is obtained using the Fresnel approximation \cite{Fresnel, PolarCS}.
	
	{Stacking all transmit steering vectors and received steering vectors as $\mathbf{A}_{\mathrm{T}} = [\mathbf{a}_{\mathrm{T}}(\psi_1), \mathbf{a}_{\mathrm{T}}(\psi_2), \cdots, \mathbf{a}_{\mathrm{T}}(\psi_L)] \in \mathbb{C}^{N_{\mathrm{T}} \times L}$ and ${\mathbf{A}}_{\mathrm{R}} = [\mathbf{s}_1 \odot {\mathbf{a}}_{\mathrm{R}}(\vartheta_1, r_1), \mathbf{s}_2 \odot {\mathbf{a}}_{\mathrm{R}}(\vartheta_2, r_2), \cdots, \mathbf{s}_L \odot {\mathbf{a}}_{\mathrm{R}}(\vartheta_L, r_L)] \in \mathbb{C}^{N_{\mathrm{R}} \times L}$, the channel in (\ref{Channel}) can be compactly expressed as 
	\begin{equation}
		\mathbf{H} = \sqrt{N_\mathrm{T}N_\mathrm{R}}{\mathbf{A}}_{\mathrm{R}} \mathbf{G} \mathbf{A}_{\mathrm{T}}^{\mathrm{H}},
		\label{Channel2}
	\end{equation}
	where $\mathbf{G} = \mathrm{diag}(g_1, \cdots, g_{L})$$\in$ $\mathbb{C}^{L \times L}$ is the path gain matrix.}
	\section{Proposed Two-Phase Estimation Scheme}
	\label{section3}
    {Due to the block-fading characteristics of wireless channels, designers often assume that the channel parameters remain constant within each coherence block. 
    Assume each coherence block can be divided into several consecutive subframes, and that $P$ consecutive subframes are used for channel estimation, as illustrated in Fig.~\ref{protocol}. Furthermore, assume each subframe can be divided into $K$ time slots.  
    In each time slot, the user activates only a single RF chain to transmit the pilot signal on one beam, while the BS utilizes all its RF chains to combine the received pilot signals from different beams.}
	
	{Given that different propagation paths exhibit distinctive SnS properties, the concurrent estimation of multipath components presents a significant challenge.} To address this, we propose a two-phase estimation scheme.
	In the first phase, the AoDs on the user side are initially estimated. Based on the obtained AoDs, the user can then select a suitable transmit beam aligned with these AoDs, allowing the BS to effectively decouple the received signals from the multipath components.
	In the second phase, the remaining path parameters are estimated. Finally, by superposing all the paths, the complete channel information is obtained. In the following sections, we will elaborate on the details of this two-phase estimation scheme.
	\vspace{-1em}
	\subsection{Phase I: AoDs Estimation}
	In the first phase, as the channel parameters are unknown, both the user and BS randomly select transmit and combiner beams from predefined codebooks. 
	{Specifically, transmit beam $\mathbf{f}_p \in \mathbb{C}^{N_{\mathrm{T}} \times 1}$ in the $p$-th subframe is selected randomly from an $N_{\mathrm{T}}$-dimension DFT codebook. The analog combining matrix $\mathbf{W}_{k} \in \mathbb{C}^{N_{\mathrm{R}} \times N_{\mathrm{RF}}}$ for the $k$-th time slot in the $p$-th subframe adheres to a constant modulus constraint, with entries independently generated from the set $\{-1/\sqrt{N_{\mathrm{R}}},1/\sqrt{N_{\mathrm{R}}}\}$.}
	
	\begin{figure}
		\centering
		\includegraphics[width=0.4\textwidth]{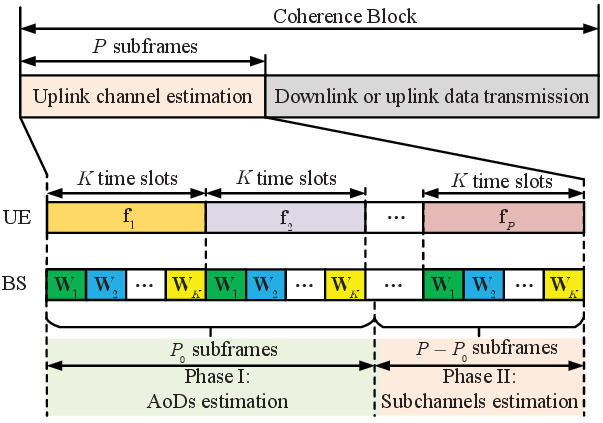}
		\caption{Two-phase uplink channel estimation scheme.}
		\label{protocol}
	\end{figure} 
	
	In this manner, the received pilot signal $\mathbf{y}_{p,k} \in \mathbb{C}^{N_{\mathrm{RF}}\times 1}$ at the BS in the $k$-th time slot of the $p$-th subframe is given by
	\begin{equation}
		\mathbf{y}_{p,k} = \mathbf{W}_{k}^{\mathrm{H}}\mathbf{H}\mathbf{f}_p\alpha_{p,k} + \mathbf{W}_{k}^{\mathrm{H}}\mathbf{n}_{p,k},
	\end{equation}
	where $\alpha_{p,k}\in \mathbb{C}$ denotes the transmit pilot symbol;
	$\mathbf{n}_{p,k} \in \mathbb{C}^{N_{\mathrm{R}} \times 1} $ is the additional white Gaussian (AWGN) noise vector, where each element obeys circularly symmetric complex Gaussian distribution with variance $\sigma^2_{\mathrm{N}}$.
	Without loss of generality, we assume $\alpha_{p,k} = 1$ for all $p$ and $k$, since the pilot sequence is known at the BS and can be readily eliminated.
	By collecting the received pilot signals from different time slots corresponding to the $p$-th transmit beam, the pilot signal $\mathbf{y}_{p} = [\mathbf{y}_{p,1}^{\mathrm{T}}, \mathbf{y}_{p,2}^{\mathrm{T}}, \cdots, \mathbf{y}_{p,K}^{\mathrm{T}}]^{\mathrm{T}} \in \mathbb{C}^{M\times 1}$ with $M = KN_{\mathrm{RF}}$ in the $p$-th subframe can be expressed as
	\begin{equation}
		\mathbf{y}_{p} = \mathbf{W}^{\mathrm{H}}\mathbf{H}\mathbf{f}_p + \overline{\mathbf{n}}_{p},
	\end{equation}
	where $\mathbf{W} = [\mathbf{W}_{1}^{\mathrm{T}}, \mathbf{W}_{2}^{\mathrm{T}}, \cdots, \mathbf{W}_{K}^{\mathrm{T}}] \in \mathbb{C}^{N_{\mathrm{R}} \times M}$; $\overline{\mathbf{n}}_{p} = [(\mathbf{W}_1 \mathbf{n}_{p,1})^\mathrm{T}, (\mathbf{W}_2 \mathbf{n}_{p,2})^\mathrm{T}, \cdots, (\mathbf{W}_{K} \mathbf{n}_{p, K})^\mathrm{T}]^{\mathrm{T}} \in \mathbb{C}^{M \times 1}$ is the equivalent noise vector with covariance matrix
	\begin{equation}
		\mathbf{R} = \mathrm{blkdiag}(\sigma^2_{\mathrm{N}} \mathbf{W}_1^{\mathrm{H}}\mathbf{W}_1, \cdots, \sigma^2_{\mathrm{N}} \mathbf{W}_{K}^{\mathrm{H}}\mathbf{W}_{K}).
		\label{Rc}
	\end{equation}
	
	Assuming the utilization of $P_0$ subframes in the first phase\footnote{{To balance pilot overhead and estimation performance, we consider to set $P_0 = 3L$ for all subsequent simulations. {Notably, in practice, prior knowledge of $L$ (the number of significant paths) can often be determined through long-term, site-specific measurements, model-order selection algorithms, or by employing compressive sensing (CS)-based support recovery algorithms.}}}, thus, we have
	\begin{equation}
		\mathbf{Y}_0 = \mathbf{W}^{\mathrm{H}}\mathbf{H}\mathbf{F} + \overline{\mathbf{N}},
		\label{Pilot}
	\end{equation}
	where $\mathbf{F} = [\mathbf{f}_1, \cdots, \mathbf{f}_{P_0}] \in \mathbb{C}^{N_{\mathrm{T}}\times P_0}$,  $\mathbf{Y}_0 = [\mathbf{y}_1, \cdots, \mathbf{y}_{P_0}] \in \mathbb{C}^{M\times P_0}$ and $\overline{\mathbf{N}} = [\overline{\mathbf{n}}_1, \cdots, \overline{\mathbf{n}}_{P_0}] \in \mathbb{C}^{M\times P_0}$ are the stacked transmit beamforming matrix, received pilot matrix and noise matrix, respectively. Utilizing (\ref{Channel2}), (\ref{Pilot}) can be rewritten as $\mathbf{Y}_0 =\mathbf{C}\mathbf{A}_{\mathrm{T}}^{\mathrm{H}}\mathbf{F} + \overline{\mathbf{N}}$,
    where $\mathbf{C} =\sqrt{N_\mathrm{T}N_\mathrm{R}} \mathbf{W}^{\mathrm{H}}{\mathbf{A}}_{\mathrm{R}} \mathbf{G}$. Define $\mathbf{M} = \mathbf{A}_{\mathrm{T}} \mathbf{C}^{\mathrm{H}}$, thus, the recovery of $\psi_l$ based on $\mathbf{Y}_0^{\mathrm{H}}$ can be formulated as a semidefinite program (SDP) problem \cite{atomic}, i.e., 
    \begin{equation}
    	\begin{aligned}
    		\min_{{\mathbf{u}}, \, {\mathbf{M}}, \, {\mathbf{Z}}} \quad &\frac{\mu}{2M} \mathrm{tr}(\mathbf{Z}) + \frac{\mu}{2N_{\mathrm{T}}} \mathrm{tr}\left(\mathrm{Toep}(\mathbf{u})\right)\\
    		&+ \frac{1}{2} \left\|\mathbf{Y}_0^{\mathrm{H}}-\mathbf{F}^{\mathrm{H}}\mathbf{M}\right\|^2_\mathrm{F}\\
    		& \mathrm{s.t.} \begin{bmatrix}
    			\mathrm{Toep}(\mathbf{u}) & \mathbf{M}\\
    			\mathbf{M}^{\mathrm{H}} & \mathbf{Z}
    		\end{bmatrix} \succeq 0,
    	\end{aligned}
    	\label{SDP}
    \end{equation}
	where $\mu$ is a regularization factor, $\mathbf{u}$ and $\mathbf{Z}$ are two auxiliary variables; {$\mathrm{Toep}(\mathbf{u})$ is a symmetric Toeplitz matrix with $\mathbf{u}$ being its first row.}
	Notably, the problem in (\ref{SDP}) is convex and can be directly solved using CVX~\cite{SDP_CVX}. 
	The recovery of $\hat{\psi}_l$ is then based on the solution
	of $\mathrm{Toep}(\mathbf{u})$ by the Multiple Signal Classification (MUSIC) or Estimating Signal Parameter via Rotational Invariance Techniques algorithms \cite{music,esprit}.
	
	{In the first stage, the computational complexity of AoD estimation depends on the size of the positive semi-definite matrix in (\ref{SDP}), which is $\mathcal{O}((N_{\mathrm{T}} + M)^{3.5})$. Notably, in mmWave MIMO systems, the angle information $\psi_l$ changes more slowly compared to the variations in the fading coefficients $g_l$ \cite{AoD_Statistical}. Therefore, AoD estimation does not need to be performed within every coherence block. Instead, it is only required when the angle information changes. By averaging the computational complexity of AoD estimation over multiple coherent blocks, it becomes more manageable and acceptable.}
	
	\subsection{Phase II: Subchannels Estimation}
	Once the AoD parameters are obtained, we can appropriately design the transmit beams in the subsequent subframes to align with the estimated AoDs. In particular, we have $\mathbf{f}_p = \mathbf{a}_{\mathrm{T}}(\hat{\psi}_l)$ for $P_0+1 \le p \le P$ and $1\le l \le L$ with $P-P_0 \ge L$. In this manner, the received signal in the $(P_0+l)$-th subframe can be expressed as
	\begin{equation}
		\begin{aligned}
			\mathbf{y}_{P_0+l} = \mathbf{W}^{\mathrm{H}}\mathbf{H}
			\mathbf{a}_{\mathrm{T}}(\hat{\psi}_l) + \overline{\mathbf{n}}_{P_0+l}.
		\end{aligned}
		\label{y_pl}
	\end{equation}
	Due to the asymptotic orthogonality of the transmit steering vector, we have $\mathbf{a}_{\mathrm{T}}^{\mathrm{H}}(\psi_l)\mathbf{a}_{\mathrm{T}}(\hat{\psi}_l) \approx 1$, if $\hat{\psi}_l = {\psi}_l$, otherwise 0.
	Therefore, (\ref{y_pl}) can be rewritten as 
	\begin{equation}
		\mathbf{y}_{P_0+l} \approx \mathbf{W}^{\mathrm{H}} \mathbf{h}_l + \overline{\mathbf{n}}_{P_0+l},
		\label{pilot3}
	\end{equation}
	{where $\mathbf{h}_l \triangleq \sqrt{N_\mathrm{T}N_\mathrm{R}} g_l \mathbf{S}_l \mathbf{a}_R(\vartheta_l, r_l)$ is defined as the $l$-th subchannel corresponding to the $l$-th path with $\mathbf{S}_l = \mathrm{diag}(\mathbf{s}_l)$}. For notation simplification, in the subsequent discussion, we simplify $\mathbf{y}_{P_0+l}$ and $\overline{\mathbf{n}}_{P_0+l}$ as $\mathbf{y}_{l}$ and $\overline{\mathbf{n}}_{l}$.
	
	{In the second phase, our aim is to estimate the subchannel $\mathbf{h}_l$ and extract the corresponding visibility indicator vector $\mathbf{s}_l$ according to the received signal $\mathbf{y}_l$. To enable efficient estimation, it is crucial to leverage the structured sparsity of channels. From the perspective of sparsity level, the polar domain emerges as the preferred choice for NF channels \cite{PolarCS, NFBT1, NFBT2}. However, it necessitates sampling in both the distance and angle domains, resulting in a significantly larger codebook size and increased computational complexity. Consequently, we propose to utilize the angular-domain (wavenumber-domain) sparsity as an alternative, as demonstrated in our previous works \cite{LoS_Angular,Turbo-OMP}.}
	{Specifically, through Fourier plane wave decomposition, the SnS subchannel can be approximated as
	\begin{equation}
		\mathbf{h}_l = \mathbf{D}\mathbf{c}_l \overset{(a)}{=} \mathbf{S}_l\mathbf{D}\mathbf{c}_l,
		\label{IDFT}
	\end{equation}
	where $\mathbf{D} \in \mathbb{C}^{N_{\mathrm{R}}\times Q}$ denotes the angular-domain transformation matrix, as constructed in \cite[Eq. (5)]{Turbo-OMP}, and $\mathbf{c}_l \in \mathbb{C}^{Q\times1}$ represents the angular channel vector. 
	The equality in $(a)$ holds because $s_{l,n} \in \{0, 1\}$, and $\mathbf{h}_l$ and $\mathbf{s}_l$ share the same positions of non-zero elements.
	Moreover, due to the limited array size, only a few angular components in $\mathbf{c}_l$ significantly contribute to the SnS subchannel, and these significant components are concentrated within a specific spatial frequency range. In other words, $\mathbf{c}_l$ exhibits block-sparsity.}
	
	\begin{remark}
		{Notably, although the SnS properties are implicitly captured in $\mathbf{c}_l$, accurately extracting SnS information directly from $\mathbf{c}_l$ is challenging. An alternative approach is to reconstruct the subchannel $\hat{\mathbf{h}}_l$ based on the estimated $\hat{\mathbf{c}}_l$, and then obtain SnS information using energy detection-based methods \cite{HanYu}. However, these methods depend on the reconstruction accuracy of $\hat{\mathbf{h}}_l$ and are sensitive to noise levels. In particular, under low SNR conditions, this approach may suffer from significant performance degradation. To address this issue, we propose explicitly characterizing the SnS properties, as shown in (\ref{IDFT}), which allows for direct estimation of $\mathbf{s}_l$.}
	\end{remark}
	Utilizing the angular representation in (\ref{IDFT}), the received pilot signal in (\ref{pilot3}) can be rewritten as 
	\begin{equation}
		\mathbf{y}_l = \mathbf{W}^{\mathrm{H}} \mathbf{S}_l\mathbf{D}\mathbf{c}_l + \overline{\mathbf{n}}_{l}.
		\label{angular_channel}
	\end{equation} 
	{According to (\ref{angular_channel}), the estimation of subchannel $\mathbf{h}_l$ is transformed into the joint estimation of $\mathbf{S}_l$ and $\mathbf{c}_l$.}
	Since the equivalent noise $\overline{\mathbf{n}}_{l}$ is colored, we consider to perform a pre-whitening procedure. {Assume that the noise covariance matrix $\mathbf{R}$ in (\ref{Rc}) can be decomposed by the Cholesky factorization as $\mathbf{R} =
	\mathbf{B}\mathbf{B}^{\mathrm{H}}$}, where $\mathbf{B} \in \mathbb{C}^{M\times M}$ is a lower triangular matrix. In this manner, we can obtain the pre-whitening transformation matrix $\mathbf{B}^{-1}$. Then, the whitened received pilot signal from the $l$-th path can be rewritten as
	\begin{equation}
		\widetilde{\mathbf{y}}_l = \mathbf{B}^{-1}\mathbf{y}_l = \mathbf{P}\mathbf{S}_l\mathbf{D}\mathbf{c}_l +  \widetilde{\mathbf{n}}_l = \boldsymbol{\Phi}_l\mathbf{c}_l +  \widetilde{\mathbf{n}}_l,
		\label{pilot4}
	\end{equation}
	where $\mathbf{P} = \mathbf{B}^{-1}\mathbf{W}^{\mathrm{H}}$, $\widetilde{\mathbf{n}}_l = \mathbf{B}^{-1}\overline{\mathbf{n}}_{l}$ and $\boldsymbol{\Phi}_l = \mathbf{P}\mathbf{S}_l\mathbf{D}$. After whitening, the covariance matrix of the noise $\widetilde{\mathbf{n}}_l$ is $\mathbf{B}^{-1}\mathbf{R}\mathbf{B}^{-\mathrm{H}} = \beta^{-1} \mathbf{I}_{M\times M}$, where $\beta$ denotes the noise precision parameter.
	
	{From (\ref{pilot4}), it might seem intuitive to estimate $\boldsymbol{\Phi}_l$ and $\mathbf{c}_l$ using bilinear inference methods \cite{Bi_GAMP}. However, existing approaches encounter significant challenges. Specifically, the dependencies among $\mathbf{P}$, $\mathbf{S}_l$, and $\mathbf{D}$ complicate the establishment of a statistical prior model for $\boldsymbol{\Phi}_l$, which restricts the direct applicability of bilinear generalized approximate message passing (Bi-GAMP) algorithms.} This highlights the pressing need for novel channel estimation techniques capable of addressing the challenges posed by the SnS property in XL-MIMO systems.
	\section{Proposed Subchannel Estimation Algorithm}
	\label{section4}
	In this section, we formulate the subchannel estimation as a layered Bayesian inference problem, and propose a computationally effective algorithm based on variational message passing and belief propagation.
	\vspace{-1em}
	\subsection{Probability Model and Factor Graph Representations}
	{In the context of our work, the measurement matrix $\mathbf{P}$ depends on the whitening matrix, which may lead to an ill-conditioned measurement matrix. To mitigate the potential divergence problems associated with the generic measurement matrix $\mathbf{P}$, we first perform a unitary transformation \cite{GAMP2, GAMP10} on the received pilot signal, i.e.,}
	\begin{equation}
		\mathbf{r}_l = \mathbf{A}\mathbf{S}_l\mathbf{D}\mathbf{c}_l + \mathbf{n}_l,
		\label{model}
	\end{equation}
	where $\mathbf{r}_l = \mathbf{U}^{\mathrm{H}}\widetilde{\mathbf{y}}_l$, $\mathbf{A} = \boldsymbol{\Lambda}\mathbf{V}^\mathrm{H}$, $\mathbf{n}_l = \mathbf{U}^{\mathrm{H}}\widetilde{\mathbf{n}}_l$. Here, $\mathbf{U}$, $\boldsymbol{\Lambda}$, and $\mathbf{V}$ are obtained through the singular value decomposition (SVD) of the measurement matrix $\mathbf{P}$, i.e., we have $\mathbf{P} = \mathbf{U}\boldsymbol{\Lambda}\mathbf{V}^\mathrm{H}$.
	Note that, since $\mathbf{U}$ is a unitary matrix,
	$\mathbf{n}_l$ is still a zero mean Gaussian noise vector with the same covariance matrix as $\widetilde{\mathbf{n}}_l$.
	
	{Although the estimation problem of $\mathbf{s}_l$ and $\mathbf{c}_l$ in (\ref{model}) can be categorized as a parametric bilinear estimation problem, the prior probability model in \cite{Para_Bi_Est} does not consider the SnS XL-MIMO channel characteristics. Specifically, the block-sparsity of $\mathbf{c}_l$ and spatial correlations among $s_{n}$ are not fully incorporated. To address these issues, we propose to utilize the hierarchical prior and Markov-chain based prior models to characterize the angular channel $\mathbf{c}_l$ and visibility indicator vector $\mathbf{s}_l$, respectively.} In particular, to capture the angular-domain block sparsity, we assume that the widely used two-layer hierarchical prior for $\mathbf{c}_l$ \cite{GAMP3}, where each element ${c}_{l,q}$ has a conditionally independent distribution expressed as
	\begin{equation}
		p(\mathbf{c}_{l}|\boldsymbol{\gamma}_{l}) = \prod_{q=1}^{Q}p({c}_{l,q}|{\gamma}_{l,q})=\prod_{q=1}^{Q}\mathcal{CN}(c_{l,q}; 0, \gamma^{-1}_{l,q}),
		\label{prior_c}
	\end{equation} 
	where {$\gamma_{l,q} \sim \mathcal{G}a(\gamma_{l,q};\xi, \eta)$} is modeled as a Gamma prior with $\xi$ and $\eta$ as the shape parameters to ensure a positive variance. It is evident that $c_{l,q}$ converges to zero as $\gamma_{l,q}^{-1}$ approaches zero, capturing the angular-domain sparsity.
	
	In terms of visibility indicator vector $\mathbf{s}_l$, since the VR exhibits spatial-correlated properties, the non-zero elements of $\mathbf{s}_{l}$ usually concentrate on a specific subset of the whole array. 
	To capture the spatial correlation, the prior distribution of visibility indicator vector $\mathbf{s}_l$ can be characterized with a one-order Markov chain as 
	\begin{equation}
		p(\mathbf{s}_{l}) = \prod_{n=1}^N p(s_{l,n}|s_{l,n-1}),
		\label{prior_s}
	\end{equation}
	where $p(s_{l,1}|s_{l,0}) = (1-\phi_l)\delta(1-s_{l,1})+ \phi_l\delta(s_{l,1})$, where $\phi_l$ reflects the sparsity level of SnS channels. The transition probability of Markov chain is given by
	\begin{equation}
		p(s_{l,n}|s_{l,n-1}) = \left\{
		\begin{aligned}
			(1-p_{01})^{1-s_{l,n}}p_{01}^{s_{l,n}}&,s_{l,n-1} = 0, \\
			p_{10}^{1-s_{l,n}} (1-p_{10})^{s_{l,n}}&,s_{l,n-1} = 1,\\
		\end{aligned}
		\right.
	\end{equation}
	where $p_{01} = p(s_{l,n} = 0|s_{l,n-1} = 1)$ and other three transition probabilities are defined similarly. By the steady-state
	assumption, the Markov chain can be completely characterized
	by two parameters $\phi_l$ and $p_{10}$. The other three transition
	probabilities can be easily obtained as $p_{01} = \phi_l p_{10}/(1-\phi_l)$, $p_{00} = 1-p_{01}$, and $p_{11} = 1-p_{10}$, respectively.
	
	\begin{figure}
		\centering
		\includegraphics[width=0.45\textwidth]{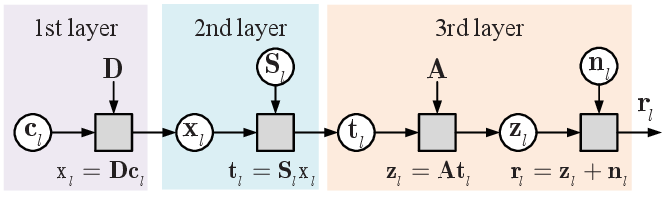}
		\caption{Proposed three-layer Bayesian inference scheme.}
		\label{Three-layer}
	\end{figure} 
	
		\begin{table}
		\centering
		\caption{Factor and Distribution in (\ref{joint_pro1})}
		\setlength{\tabcolsep}{4mm}{
			\begin{tabular}{c c c}
				\toprule
				\makecell[c]{Factor} & \makecell[c]{Distribution} &  \makecell[c]{Function} \\
				\midrule
				$f_\beta(\beta)$ &$p(\beta)$  & \makecell[c]{$\beta^{-1}$}  \\
				
				$f_{\boldsymbol{\gamma}_l} (\boldsymbol{\gamma}_l)$& $p(\boldsymbol{\gamma}_l)$  & \makecell[c]{$\mathcal{G}a(\gamma_{l,q};\xi, \eta)$} \\
				
				$f_{\mathbf{c}_l}(\mathbf{c}_l, \boldsymbol{\gamma}_l)$& $p(\mathbf{c}_l|\boldsymbol{\gamma}_l)$ & (\ref{prior_c}) \\
				
				$f_{\mathbf{s}_l}(\boldsymbol{s}_l)$& $p(\mathbf{s}_l)$  & \makecell[c]{(\ref{prior_s})}  \\
				
				$f_{\mathbf{x}_l}(\mathbf{x}_l,\mathbf{c}_l)$& $p(\mathbf{x}_l|\mathbf{c}_l)$ &$\delta(\mathbf{x}_l -\mathbf{D}\mathbf{c}_l)$  \\
				
				$f_{\mathbf{t}_l}(\mathbf{t}_l, \mathbf{s}_l, \mathbf{x}_l)$ &$p(\mathbf{t}_l|\mathbf{s}_l, \mathbf{x}_l)$ &$\delta(\mathbf{t}_l - \mathbf{S}_l \mathbf{x}_l)$  \\
				
				$f_{\mathbf{z}_l}(\mathbf{z}_l, \mathbf{t}_l)$& $p(\mathbf{z}_l|\mathbf{t}_l)$ &$\delta(\mathbf{z}_l - \mathbf{A}\mathbf{t}_l)$  \\
				
				$f_{\mathbf{r}_l}(\mathbf{r}_l, \mathbf{z}_l, \beta)$& $p(\mathbf{r}_l|\mathbf{z}_l,\beta)$ &$\mathcal{CN}(\mathbf{r}_l; \mathbf{z}_l, \beta^{-1}\mathbf{I}_M)$  \\
				\bottomrule
			\end{tabular}
		}
		\label{proba}
	\end{table}
	
	{Due to the introduction of the Markov-chain-based prior model $p(\mathbf{s}_l)$, the parametric bilinear estimation scheme \cite{Para_Bi_Est} may become less applicable. Specifically, the Markov-chain structure introduces a complex graph topology with loops among the nodes $s_{l,n}$, $p(s_{l,n} | s_{l,n-1})$, and $p(r_{l,m}| \mathbf{s}_l, \mathbf{c}_l)$, leading to difficulties in performing forward and backward message updates.
	To address this issue, we reformulate the SnS channel estimation problem as a layered Bayesian inference scheme. In this approach, $\mathbf{s}_l$ and $\mathbf{c}_l$ are estimated simultaneously across different layers, effectively circumventing the complex graph topology between variable and factor nodes.
	The layered Bayesian inference scheme is illustrated in Fig.~\ref{Three-layer}, where blank circles represent the realizations of random variables to be estimated, and gray boxes indicate the corresponding functional operations, with $\mathbf{x}_{l} \triangleq \mathbf{D}\mathbf{c}_l$, $\mathbf{t}_l \triangleq \mathbf{S}_l\mathbf{x}_l$, and $\mathbf{z}_l \triangleq \mathbf{A}\mathbf{t}_l$.}
	{Based on the prior distribution of $\mathbf{c}_l$ and $\mathbf{s}_l$ and the proposed layered inference scheme}, the joint posterior distribution of $\boldsymbol{\gamma}_l$, $\mathbf{c}_l$, $\mathbf{x}_l$, $\mathbf{s}_l$, $\mathbf{t}_l$, $\mathbf{z}_l$, and $\beta$ given $\mathbf{r}_l$ can be factorized as (\ref{joint_pro1}), as shown in the top of next page,
	where the involved probability distributions are listed in Table~\ref{proba}. 
	For brevity, we omit the subscript $l$ in the subsequent derivation.
	Moreover, the factor graph representation of (\ref{joint_pro1}) is depicted in Fig.~\ref{GF}, where the gray squares represent the factor nodes, and the blank circles represent the variable nodes.
	
	\begin{figure*}
		\centering
		\includegraphics[width=0.7\textwidth]{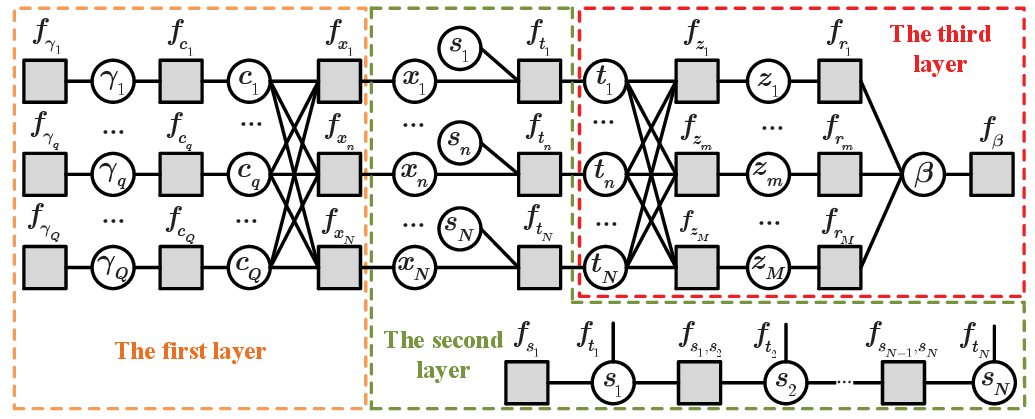}
		\caption{Illustration of factor graph representations of (\ref{joint_pro1}).}
		\label{GF}
	\end{figure*}

	\begin{figure*}[!t]
		\normalsize
		\setcounter{MYtempeqncnt}{\value{equation}}
		\setcounter{equation}{19}
		\begin{equation}
			\begin{aligned}
				&p(\boldsymbol{\gamma}_l, \mathbf{c}_l, \mathbf{x}_l, \mathbf{s}_l, \mathbf{t}_l, \mathbf{z}_l, \beta| \mathbf{r}_l)
				\propto p(\mathbf{r}_l|\mathbf{z}_l,\beta)p(\beta)p(\mathbf{z}_l|\mathbf{t}_l)p(\mathbf{t}_l|\mathbf{s}_l,\mathbf{x}_l) p(\mathbf{s}_l) p(\mathbf{x}_l|\mathbf{c}_l) p(\mathbf{c}_l|\boldsymbol{\gamma}_l) p(\boldsymbol{\gamma}_l) \\ 
				\triangleq& 
				\underbrace{f_{\mathbf{x}_l}(\mathbf{x}_l,\mathbf{c}_l)f_{\mathbf{c}_l}(\mathbf{c}_l,\boldsymbol{\gamma}_l)  f_{\boldsymbol{\gamma}_l}(\boldsymbol{\gamma}_l) }_{\text{the first layer}}
				\underbrace{f_{\mathbf{t}_l}(\mathbf{t}_l,\mathbf{s}_l,\mathbf{x}_l)
					f_{\mathbf{s}_l}(\mathbf{s}_l)}_{\text{the second layer}}
				\underbrace{f_{\mathbf{r}_l}(\mathbf{r}_l,\mathbf{z}_l,\beta) f_{\mathbf{z}_l}(\mathbf{z}_l,\mathbf{t}_l)f_{\beta}(\beta)}_{\text{the third layer}},
			\end{aligned}
			\label{joint_pro1}
		\end{equation}
		\hrulefill
		\vspace*{4pt}
	\end{figure*}

	{Utilizing the probabilistic model provided in (\ref{joint_pro1})}, the optimal estimators of $\mathbf{s}$ and $\mathbf{c}$ under the minimum mean square error (MMSE) principle can be derived as
	\begin{align}
		\label{MMSE1}\hat{{s}}_{n} &= \int {s}_{n} p(\boldsymbol{\gamma}, \mathbf{c}, \mathbf{x}, \mathbf{s}, \mathbf{t}, \mathbf{z}, \beta| \mathbf{r}) \mathrm{d}\boldsymbol{\gamma}\, \mathrm{d}\mathbf{c} \, \mathrm{d}\mathbf{t} \, \mathrm{d}\mathbf{z}\, \mathrm{d}\beta \, \mathrm{d}\mathbf{s},\\
		\label{MMSE2}\hat{{c}}_{q} &= \int {c}_{q} p(\boldsymbol{\gamma}, \mathbf{c}, \mathbf{x}, \mathbf{s}, \mathbf{t}, \mathbf{z}, \beta| \mathbf{r}) \mathrm{d}\boldsymbol{\gamma}\, \mathrm{d}\mathbf{s} \, \mathrm{d}\mathbf{t} \, \mathrm{d}\mathbf{z}\, \mathrm{d}\beta \, \mathrm{d}\mathbf{c}.
	\end{align}
	Given the substantial number of antenna elements in XL-MIMO systems, both MMSE estimators (\ref{MMSE1}) and (\ref{MMSE2}) involve high-dimensional integrals, making them impractical to compute. 
	Recently, low-complexity techniques such as GAMP \cite{MP_Contributions,Multi-layer} have been extensively employed for solving the maximum a posterior (MAP) estimation problem.
	Motivated by the success of the multi-layer GAMP in {cascaded estimation problems \cite{Multi-layer}}, we propose a computationally efficient TL-GAMP algorithm to solve the MMSE estimation problem in (\ref{MMSE1}) and (\ref{MMSE2}), where $\hat{{s}}_{n}$ and $\hat{{c}}_{q}$ are estimated in separate layers. 
	In the following, we will elaborate on the details of message passing in each layer.
	\vspace{-1em}
	\subsection{Message Passing in the First Layer}
	\begin{figure}
		\centering
		\includegraphics[width=0.45\textwidth]{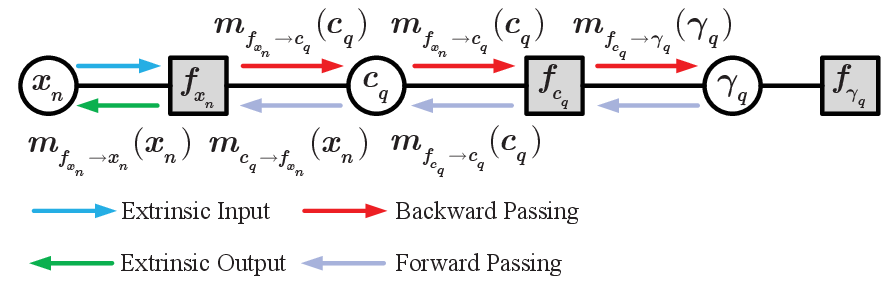}
		\caption{Illustration of message passing in the first layer.}
		\label{l1}
	\end{figure} 
	In the first layer, the sparsity of angular the channel $\mathbf{c}$ is captured, and corresponding coefficients ${c}_q$ are estimated by exploiting the prior distribution information and the likelihood messages from the second layer. Meanwhile, the prior distribution information associated with $\mathbf{x}$ is output for the second layer. Fig. \ref{l1} illustrates the message passing in the first layer.
	
	{Based on the Central Limit Theorem (CLT), we denote the message from $x_n$ to $f_{x_n}$ as $m_{x_n \rightarrow f_{x_n}}(x_n) \propto \mathcal{CN}(x_n; \vec{x}_n, \vec{\nu}_{n}^{x})$ defined below (\ref{m_ft2x}), and the belief of $c_q$ as $b(c_q) \propto \mathcal{CN}(c_q; \hat{c}_q, \hat{\nu}_q^{c})$ defined in (\ref{bcq}), respectively. Then, according to the variational message passing rules \cite{GAMP3,GAMP4}, the message from $f_{c_q}$ to $\gamma_q$ can be derived as }
	\begin{equation}
		\begin{aligned}
			m_{f_{c_q} \rightarrow \gamma_q}(\gamma_q) &\propto \exp\left\{\int \ln f_{c_q}(c_q, \gamma_q) b(c_q) \mathrm{d}c_q \right\}\\
			 &\propto \gamma_q\exp\left\{-\gamma_q \left(\left|\hat{c}_q\right|^2 + \hat{\nu}^c_q\right)\right\}.
		\end{aligned}
		\label{m_fc2ga}
	\end{equation}
	
	Combining the likelihood  message $m_{f_{c_q} \rightarrow \gamma_q}(\gamma_q)$ in (\ref{m_fc2ga}) and the prior message $f_{\gamma_q}(\gamma_q)$ associated with $\gamma_q$, the belief of $\gamma_q$ is denoted as
	\begin{equation}
		\begin{aligned}
			b(\gamma_q) &\propto m_{f_{c_q} \rightarrow \gamma_q}(\gamma_q) f_{\gamma_q}(\gamma_q)\\
			&= \gamma_q^{\xi}\exp\left\{-\gamma_q \left(\eta + \left|\hat{c}_q\right|^2 + \hat{\nu}^c_q \right)\right\}.
		\end{aligned} 	
		\label{b_ga}
	\end{equation}
	From (\ref{b_ga}), $b(\gamma_q)$ obeys the Gamma distribution with the shape parameters $\xi + 1$ and $\eta + \left|\hat{c}_q\right|^2 + \hat{\nu}^c_q$. Hence, the approximate posterior mean of ${\gamma}_q$ is given by
	\begin{equation}
		\hat{\gamma}_q = \frac{\xi+1}{\eta + \left|\hat{c}_q\right|^2 + \hat{\nu}^c_q}.
		\label{gamma}
	\end{equation}
	With the belief $\gamma_q$ in (\ref{b_ga}) and variational message passing rule, the message from $f_{c_q}$ to $c_q$ is given by 
	\begin{equation}
		\begin{aligned}
			m_{f_{c_q} \rightarrow c_q}(c_q) &\propto \exp\left\{\int \ln f_{c_q}(c_q, \gamma_q) b(\gamma_q) \mathrm{d}\gamma_q \right\}\\
			&= \mathcal{CN}(c_q; 0, \hat{\gamma}_q^{-1}).
		\end{aligned}
		\label{m_fga2c}
	\end{equation}
	
	Combining $m_{f_{c_q} \rightarrow c_q}(c_q)$ in (\ref{m_fga2c}) and the message $m_{c_q \rightarrow f_{c_q}} (c_q)$ derived in (\ref{m_c2fc}), the belief of $c_q$ is given by
	\begin{equation}
		\begin{aligned}
			b(c_q) = m_{c_q \rightarrow f_{c_q}} (c_q) m_{f_{c_q} \rightarrow c_q}(c_q)
			=\mathcal{CN}(c_q; \hat{c}_q, \hat{\nu}^c_q),
		\end{aligned}
		\label{bcq}
	\end{equation}
	where $\hat{\nu}_q^c$ and $\hat{c}_q $ are respectively given by
	\begin{align}
		\label{v_cq} \hat{\nu}_q^c = \frac{\vec{\nu}^c_q}{1 + \vec{\nu}^c_q\hat{\gamma}_q}, \quad
		\hat{c}_q     = \frac{\vec{c}_q}{1+ \hat{\gamma}_q \vec{\nu}_q^c}.
	\end{align}
	
	Based on the CLT, we refer to the message from $f_{x_n}$ to $c_q$ as $m_{f_{x_n}\rightarrow c_q}(c_q) \propto \mathcal{CN}(c_q; \vec{c}_{n,q}, \vec{\nu}^c_{n,q})$ derived in (\ref{m_x2c}). According to the belief propagation rule \cite{Vector_AMP}, the message from $c_q$ to $f_{x_n}$ can be derived as
	\begin{equation}
		m_{c_q \rightarrow f_{x_n}}(c_q) \propto \frac{b(c_q)}{m_{f_{x_n}\rightarrow c_q}(c_q)}= \mathcal{CN}(c_q; \cev{c}_{q,n}, \cev{\nu}^c_{q,n}),
		\label{m_c2fx}
	\end{equation}
	where $\cev{\nu}^c_{q,n}$ and $\cev{c}_{q,n}$ are respectively given by
	\begin{equation}
		\begin{aligned}
			\label{vc_qn} \cev{\nu}^c_{q,n} = \frac{\hat{\nu}_q^c \vec{\nu}_{n,q}}{\vec{\nu}_{n,q} -\hat{\nu}_q^c}, \quad
			\cev{c}_{q,n} = \cev{\nu}^c_{q,n}\left(\frac{\hat{c}_q}{\hat{\nu}_q^c}  - \frac{\vec{c}_{n,q}}{\vec{\nu}_{n,q}^c}\right).
		\end{aligned}
	\end{equation}

	Given the message $m_{c_q \rightarrow f_{x_n}} (x_n)$ in (\ref{m_c2fx}), the message $m_{f_{x_n} \rightarrow x_n}(x_n)$ of the first layer from $f_{x_n}$ to $x_n$ is given by
	\begin{equation}
		\begin{aligned}
			m_{f_{x_n} \rightarrow x_n}(x_n) 
			&\propto \int f_{x_n}(x_n, \mathbf{c})\prod_{q=1}^Qm_{c_q \rightarrow f_{x_n}}(c_q) \, \mathrm{d}\mathbf{c}\\
			&= \mathcal{CN}(x_n; \cev{x}_n, \cev{\nu}^x_n),
		\end{aligned}
		\label{fx2x}
	\end{equation}
	where $\cev{x}_n$ and $\cev{\nu}^x_n$ are respectively calculated as
	\begin{align}
		\cev{x}_n  = \sum_{q=1}^Q d_{n,q} \cev{c}_{q,n}, \,
		\cev{\nu}^x_n  = \sum_{q=1}^Q \left|d_{n,q}\right|^2 \cev{\nu}_{q,n}^c. 
	\end{align}
	
	Combining the messages $m_{x_n \rightarrow f(x_n)}(x_n)$ from the second layer and $m_{c_i \rightarrow f_{x_n}}(c_i)$ with $i \neq q$, the message from $f_{x_n}$ to $c_q$ can be given by
	\begin{equation}
		\begin{aligned}
			m_{f_{x_n} \rightarrow c_q} (c_q) \propto& \int f_{x_n}(x_n, \mathbf{c}) m_{x_n \rightarrow f(x_n)}(x_n)\\
			 &\prod_{i \neq q}^Q m_{c_i \rightarrow f_{x_n}}(c_i)\,\mathrm{d}\mathbf{c}_{\backslash q}\, \mathrm{d}x_n\\
			=& \mathcal{CN}(c_m; \vec{c}_{n,q}, \vec{\nu}_{n,q}^c),
		\end{aligned}
		\label{m_x2c}
	\end{equation}
	where the $\vec{c}_{n,q}$ and $\vec{\nu}_{n,q}^c$ are respectively given by
	\begin{equation}
		\begin{aligned}
			\vec{c}_{n,q} & = \frac{\vec{x}_n - \cev{x}_n + d_{n,q} \cev{c}_{q,n}}{d_{n,q}},\\
			\vec{\nu}_{n,q}^c & = \frac{\vec{\nu}^x_n + \cev{\nu}^x_n- \left|d_{n,q}\right|^2\cev{\nu}^c_{q,n}}{\left|d_{n,q}\right|^2}.
		\end{aligned}
	\end{equation}
	 
	
	Then, the message from $c_q$ to $f_{c_q}$ is calculated as
	\begin{equation}
		\begin{aligned}
			m_{c_q \rightarrow f_{c_q}}(c_q) \propto \prod_{n=1}^{N_{\mathrm{R}}} m_{f_{x_n \rightarrow c_q}}(c_q) = \mathcal{CN}(c_q; \vec{c}_q, \vec{\nu}^c_q),
		\end{aligned}
		\label{m_c2fc}
	\end{equation}
	where the variance $\vec{\nu}^c_q$ and mean $\vec{c}_q$ are respectively defined as
	\begin{align}
		\label{vc_qr} \vec{\nu}^c_q = \left({\sum_{n=1}^{N_{\mathrm{R}}}\frac{1}{\vec{\nu}_{n,q}^c}}\right)^{-1},\quad
		\vec{c}_q     = \hat{\nu}^c_q\sum_{n=1}^{N_{\mathrm{R}}}\frac{\vec{c}_{n,q}}{\vec{\nu}_{n,q}^c}.
	\end{align}
	
	\begin{figure*}[!t]
		\normalsize
		\setcounter{MYtempeqncnt}{\value{equation}}
		\setcounter{equation}{36}
		\begin{equation}
			\begin{aligned}
				m_{f_{t_n} \rightarrow s_n}(s_n)
				\propto& \int f_{t_n}(t_n, s_n, x_n) m_{t_n \rightarrow f_{t_n}}(t_n) m_{x_n \rightarrow f_{x_n}}(x_n)\, \mathrm{d}x_n \, \mathrm{d}t_n\\ 
				\overset{(a)}{=}& \mathcal{CN}(0; \vec{t}_n, \vec{\nu}^t_n)\delta(s_n) + \mathcal{CN}(0; \vec{t}_n+\cev{x}_n, \vec{\nu}^t_n+ \cev{\nu}^x_n) \delta(1-s_n)\\
				{\propto}& (1-\pi^{\mathrm{out}}_n) \delta(s_n) + \pi^{\mathrm{out}}_n\delta(1-s_n),
			\end{aligned}
			\label{m_ft2s}
		\end{equation}
		\hrulefill
		\vspace*{4pt}
	\end{figure*}
	\vspace{-1em}
	\subsection{Message Passing in the Second Layer}
	\begin{figure}
		\centering
		\includegraphics[width=0.3\textwidth]{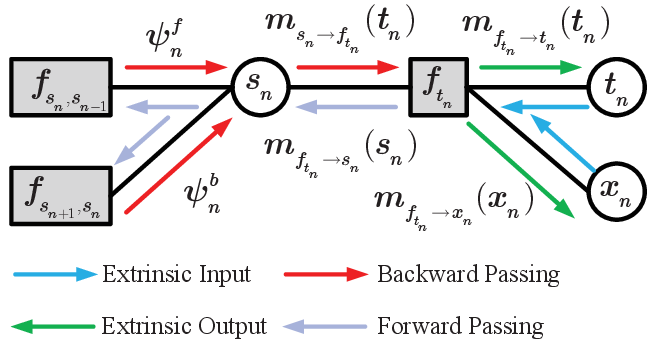}
		\caption{Illustration of message passing in the second layer.}
		\label{l2}
	\end{figure} 
	In the second layer, the SnS properties is captured, and the visibility indicator vector ${s}_n, \forall n \in \mathbb{N}_{\mathrm{R}}$ are estimated by exploiting the messages from the first and third layers. Meanwhile, the likelihood and prior information associated with $x_n$ and $t_n$ for the first and third layers are output. The message passing is shown in Fig. \ref{l2}, where $f_{s_n,s_{n-1}} \triangleq p(s_n|s_{n-1})$.
	
	Assume the message from the third layer as $m_{t_n \rightarrow f_{t_n}}(t_n) \propto \mathcal{CN}(t_n; \vec{t}_n, \vec{\nu}^t_n)$ derived in (\ref{m_t2ft2}). Thus, the message from $f_{t_n}$ to $s_n$ can be given by (\ref{m_ft2s}), as shown in the top of this page, 
	where $(a)$ is obtained according to the fact $\int \delta(x)f(x)\mathrm{d}x = f(0)$ and $s_n \in \left\{0,1\right\}$. In addition, $\pi^{\mathrm{out}}_n$ is defined as
	\begin{equation}
		\pi^{\mathrm{out}}_n = \frac{\mathcal{CN}(0; \vec{t}_n+\cev{x}_n, \vec{\nu}^t_n+ \cev{\nu}^x_n)}{\mathcal{CN}(0; \vec{t}_n, \vec{\nu}^t_n) + \mathcal{CN}(0; \vec{t}_n+\cev{x}_n, \vec{\nu}^t_n+ \cev{\nu}^x_n)}.
		\label{pi_out}
	\end{equation}
	
	With the message $m_{f_{t_n} \rightarrow s_n}(s_n)$ in (\ref{m_ft2s}), the message from $s_n$ to $f_{t_n}$ is given by
	\begin{equation}
		{m}_{s_n \rightarrow f_{t_n}}(s_n) \propto  (1-\pi^{\mathrm{in}}_n) \delta(s_{n}) + \pi^{\mathrm{in}}_n \delta(1-s_{n}),
		\label{m_s2ft}
	\end{equation} 
	where $\pi^{\mathrm{in}}_n$ is defined as 
	\begin{equation}
		\pi^{\mathrm{in}}_n = \frac{\psi^f_n \psi^b_n}{\psi^f_n \psi^b_n + (1-\psi^f_n)(1-\psi^b_n)}.
		\label{pi_in}
	\end{equation}
	Here, $\psi^f_n$ and $\psi^b_n$ are the forward and backward messages along the Markov chain, and are respectively defined as \cite{Turbo-OMP}
	\begin{equation}
		\label{psi_f} \psi^f_n = \frac{p_{01}(1-\psi^f_{n-1})(1-\pi_{n-1}^{\mathrm{out}})+ p_{11}\lambda^f_{n-1} \pi_{n-1}^{\mathrm{out}}}{(1-\psi^f_{n-1})(1-\pi_{n-1}^{\mathrm{out}})+ \psi^f_{n-1} \pi_{n-1}^{\mathrm{out}}},
	\end{equation}
	\begin{equation}
		\label{psi_b} \psi^b_n = \frac{p_{10}(1-\psi^b_{n+1})(1-\pi_{n+1}^{\mathrm{out}})+p_{11}\psi^b_{n+1}\pi_{n+1}^{\mathrm{out}}}{p_0(1-\psi^b_{n+1})(1-\pi_{n+1}^{\mathrm{out}})+p_1\psi^b_{n+1}\pi_{n+1}^{\mathrm{out}}}.
	\end{equation}
	where $p_0 = p_{10}+p_{00}$, $p_1 = p_{11}+p_{01}$.
	
	Combining the message $m_{f_{t_n} \rightarrow s_n}(s_n)$ in (\ref{m_ft2s}) and ${m}_{s_n \rightarrow f_{t_n}}(s_n)$ in (\ref{pi_in}), the belief of $s_n$ can be given by
	\begin{equation}
		b(s_n) = \frac{\psi^f_n \psi^b_n\pi_n^{\mathrm{out}}}{\psi^f_n \psi^b_n\pi_n^{\mathrm{out}} + (1-\psi^f_n)(1-\psi^b_n)(1-\pi_n^{\mathrm{out}}) }.
		\label{bsn}
	\end{equation}
	Using the obtained $b(s_n)$, $\psi^f_{0}$ in (\ref{psi_f}) and is updated in each iteration as
	\begin{equation}
		\begin{aligned}
			\psi^f_{0} = \frac{1}{N_{\mathrm{R}}} \sum_{n=1}^{N_{\mathrm{R}}}b(s_n).
		\end{aligned}
		\label{psi_0}
	\end{equation}  
	
	With the message ${m}_{s_n \rightarrow f_{t_n}}(s_n)$ in (\ref{m_s2ft}) and $m_{t_n \rightarrow f_{t_n}}(t_n)$ in (\ref{m_t2ft2}), the output message $m_{f_{t_n} \rightarrow x_n} (x_n)$ from $f_{t_n}$ to $x_n$ is given by
	\begin{equation}
		\begin{aligned}
			m_{f_{t_n} \rightarrow x_n} (x_n) \propto& \int f_{t_n}(t_n, s_n, x_n){m}_{s_n \rightarrow f_{t_n}}(s_n)\\
			&m_{t_n \rightarrow f_{t_n}}(t_n) \, \mathrm{d}s_n \, \mathrm{d}t_n\\
			\propto& \mathcal{CN}(x_n; \vec{t}_n, \vec{\nu}^t_n).
		\end{aligned}
		\label{m_ft2x}
	\end{equation}
	Furthermore, since $x_n$ is only
	connected with $f_{t_n}$ and $f_{x_n}$ in the factor graphs, the message from $x_n$ to $f_{x_n}$ also satisfies $m_{x_n \rightarrow f_{x_n}}(x_n) \propto \mathcal{CN}(x_n; \vec{t}_n, \vec{\nu}^t_n)$. 
	Similar to (\ref{m_ft2x}), with $m_{f_{x_n} \rightarrow x_n}(x_n)$ in (\ref{fx2x}) and ${m}_{s_n \rightarrow f_{t_n}}(s_n)$ in (\ref{m_s2ft}), the output message $m_{f_{t_n} \rightarrow t_n} (t_n)$ from $f_{t_n}$ to $t_n$ is given by
	\begin{equation}
		\begin{aligned}
			m_{f_{t_n} \rightarrow t_n}(t_n)
			 =(1-\pi_n^{\mathrm{in}})\delta(t_n) + \pi_n^{\mathrm{in}}\mathcal{CN}(t_n; \cev{x}_n, \cev{\nu}^x_n).
		\end{aligned}
		\label{m_ft2t}
	\end{equation}
	\vspace{-2em}
	\subsection{Message Passing in the Third Layer}
		\begin{figure}
		\centering
		\includegraphics[width=0.45\textwidth]{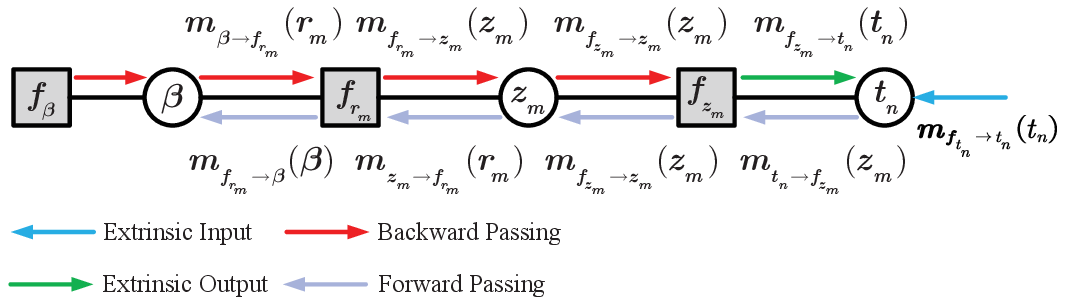}
		\caption{Illustration message passing in the third layer.}
		\label{l3}
	\end{figure} 
	In the third layer, the variables ${t}_n$, $\forall n \in \mathbb{N}_{\mathrm{R}}$, are decoupled from $\mathbf{r}$ and  estimated by combining the messages from the second layer. Meanwhile, the likelihood information associated with $t_n$ is output for the second layer. The message passing is shown in Fig. \ref{l3}.
	
	Denote the message from $f_{z_m}$ to $t_n$ and the belief of $t_n$ as $m_{f_{z_m} \rightarrow t_n} \propto \mathcal{CN}(t_n; \vec{t}_{m,n}, \vec{\nu}^t_{m,n})$ and $b(t_n) \sim \mathcal{CN}(t_n; \hat{t}_n, \hat{v}^t_n)$, respectively. According to the belief propagation rule, the message from $t_n$ to  $f_{z_m}$ can be given by 
	\begin{equation}
		\begin{aligned}
			m_{t_n \rightarrow f_{z_m}}(t_n) \propto \frac{b(t_n)}{m_{f_{z_m} \rightarrow t_n}(t_n)}= \mathcal{CN}(t_n; \cev{t}_{n,m}, \cev{\nu}^t_{n,m}),
		\end{aligned}
	\end{equation}
	where $\cev{\nu}^t_{n,m}$ and $\cev{t}_{n,m}$ are respectively given by
	\begin{equation}
		\begin{aligned}
			\cev{\nu}^t_{n,m} = \frac{\hat{\nu}^t_n \vec{\nu}^t_{m,n}}{\vec{\nu}^t_{m,n} - \hat{\nu}^t_n}, \quad
			\cev{t}_{n,m}     = \cev{\nu}^t_{n,m}\left(\frac{\hat{t}_n}{\hat{\nu}^t_n} - \frac{\vec{t}_{m,n}}{\vec{\nu}^t_{m,n}}\right).
		\end{aligned}
	\end{equation}

	With the message $m_{t_n \rightarrow f_{z_m}}(t_n)$, the message from $f_{z_m}$ to $z_m$ is given by
	\begin{equation}
		\begin{aligned}
			m_{f_{z_m} \rightarrow z_m}(z_m) &= \int f_{z_m}(z_m, \mathbf{t}) \prod_{n=1}^{N_{\mathrm{R}}} m_{t_n \rightarrow f_{z_m}}(t_n) \, \mathrm{d}\mathbf{t}\\
			&\propto \mathcal{CN}(z_m; \cev{z}_m, \cev{\nu}^z_m). 
		\end{aligned}
		\label{m_fz2z}
	\end{equation}
	$\cev{z}_m$ and $\cev{\nu}^z_m$ are respectively calculated as
	\begin{align}
		\label{z_mnl}  \cev{z}_m  = \sum_{n=1}^{N_{\mathrm{R}}}a_{m,n} \cev{t}_{n,m},
		\,
		\cev{\nu}^z_m = \sum_{n=1}^{N_{\mathrm{R}}}\left|a_{m,n}\right|^2 \cev{\nu}_{n,m}^t.
	\end{align}
	
	Denote the belief of $z_m$ as $b(z_m) \propto \mathcal{CN}(z_m; \hat{z}_m, \hat{\nu}^z_m)$ provided in (\ref{bzm}). Similar to (\ref{m_fc2ga}), the message from $f_{r_m}$ to $\beta$ is given by
	\begin{equation}
		\begin{aligned}
			m_{f_{r_m} \rightarrow \beta}(\beta) &\propto \exp\left\{\int \ln f_{r_m}(r_m, \beta) b(z_m) \mathrm{d}z_m \right\}\\
			&\propto \beta \exp\left\{-\beta \left(\left| r_m - \hat{z}_m\right|^2 + \hat{\nu}^z_m\right)\right\}.
		\end{aligned}
		\label{m_fr2beta}
	\end{equation}
	
	Combining the prior message $f_{\beta}(\beta)$ and $m_{f_{r_m} \rightarrow \beta} (\beta)$ in (\ref{m_fr2beta}), the belief of $\beta$ is given by
	\begin{equation}
		\begin{aligned}
			b(\beta) 
			\propto \beta^{M-1}\exp\left\{-\beta \sum_{m=1}^M \left(\left| r_m - \hat{z}_m\right|^2 + \hat{\nu}^z_m\right)\right\}.
		\end{aligned}
	\end{equation}
	It is observed that the belief $b(\beta)$ obeys the Gamma distribution with shape parameters $M$ and $\sum_{m=1}^M\left(\left| r_m - \hat{z}_m\right|^2 + \hat{\nu}^z_m\right)$. Thus, the approximate posterior mean $\hat{\beta}$ is given by
	\begin{equation}
		\hat{\beta} = \frac{M}{\sum_{m=1}^M\left(\left| r_m - \hat{z}_m\right|^2 + \hat{\nu}^z_m\right)}.
		\label{post_beta}
	\end{equation} 
	
	With $b(\beta)$, the message from $f_{r_m}$ to $z_m$ is given by
	\begin{equation}
		\begin{aligned}
			m_{f_{r_m} \rightarrow z_m} (z_m)
			&\propto \exp\left\{\int \ln f_{r_m}(r_m, \beta) b(\beta) \, \mathrm{d}\beta \right\}\\
			&\propto \mathcal{CN}(z_m; r_m, \hat{\beta}^{-1}). 
		\end{aligned}
		\label{m_fr2z}
	\end{equation}
	Furthermore, since $z_m$ is only
	connected with $f_{r_m}$ and $f_{z_m}$ in the factor graph, the message from $z_m$ to $f_{z_m}$ is given by $m_{z_m \rightarrow f_{z_m}}(z_m) \propto \mathcal{CN}(z_m; r_m, \hat{\beta}^{-1})$.
	Combining $m_{z_m \rightarrow f_{z_m}}(z_m)$ and $m_{f_{z_m} \rightarrow z_m}(z_m)$ in (\ref{m_fr2z}) and (\ref{m_fz2z}), the belief of $z_m$ is given by
	\begin{equation}
		\begin{aligned}
			b(z_m) \propto \mathcal{CN}(z_m; \hat{z}_m, \hat{\nu}^z_m),
		\end{aligned}
		\label{bzm}
	\end{equation}
	where $\hat{\nu}^z_m$ and $\hat{z}_m$ are respectively given by
	\begin{align}
		\label{vz_m} \hat{\nu}^z_m = \frac{\cev{\nu}^z_m}{1+\hat{\beta} \cev{\nu}^z_m}, \,
		\hat{z}_m     = \hat{\nu}^z_m\left(r_m\hat{\beta} + \frac{\cev{z}_m}{\cev{\nu}^z_m}\right).
	\end{align}
	
	With the messages $m_{z_m \rightarrow f_{z_m}}(z_m)$ and $m_{t_i \rightarrow f_{z_m}} (t_i)$ with $i \neq n$, the message from $f_{z_m}$ to $t_n$ is given by 
	\begin{equation}
		\begin{aligned}
			m_{f_{z_m} \rightarrow t_n} (t_n) =& \int f_{z_m}(z_m, \mathbf{t})m_{z_m \rightarrow f_{z_m}}(z_m)\\
			  &\prod_{i \neq n}^{N_{\mathrm{R}}} m_{t_i \rightarrow f_{z_m}}(t_i) \, \mathrm{d}z_m \, \mathrm{d}\mathbf{t}_{\backslash n}\\
			\propto& \mathcal{CN}(t_n, \vec{t}_{m,n}, \vec{\nu}^t_{m,n}),
		\end{aligned}
		\label{m_fz2t}
	\end{equation}
	where $\vec{t}_{m,n}$ and $\vec{\nu}_{m,n}^t$ are respectively given by
	\begin{equation}
		\begin{aligned}
			\label{tmn_r} \vec{t}_{m,n} &= \frac{r_m - \cev{z}_m + a_{m,n}\cev{t}_{n,m}}{a_{m,n}}, \\
			\vec{\nu}_{m,n}^t &= \frac{\hat{\beta}^{-1} + \cev{\nu}^z_m- \left|a_{m,n}\right|^2\cev{\nu}^t_{n,m}}{\left|a_{m,n}\right|^2}.
		\end{aligned}
	\end{equation}
	
	Combining the message $m_{f_{z_m} \rightarrow t_n}(t_n)$ for all $m$, the output message $m_{f_{t_n} \rightarrow t_n} (t_n)$ from $t_n$ to $f_{t_n}$ is given by 
	\begin{equation}
		m_{t_n \rightarrow f_{t_n}} (t_n) = \prod_{m=1}^M m_{f_{z_m} \rightarrow t_n} (t_n) \propto \mathcal{CN}(t_n; \vec{t}_n, \vec{\nu}^t_n),
		\label{m_t2ft2}
	\end{equation}
	where the variance $\vec{t}_n$ and mean $\vec{\nu}^t_n$ are respectively given by
	\begin{align}
		\label{vm_t2ft} \vec{\nu}^t_n = \left(\sum_{m=1}^M \frac{1}{\vec{\nu}_{m,n}^t}\right)^{-1}, \quad
		\vec{t}_n     = \vec{\nu}^t_n\sum_{m=1}^M\left(\frac{\vec{t}_{m,n}}{\vec{\nu}_{m,n}^t}\right).
	\end{align}
	
	With $m_{f_{t_n} \rightarrow t_n}(t_n)$  and  $m_{t_n \rightarrow f_{t_n}} (t_n)$, we have 
	\begin{equation}
		\begin{aligned}
			b(t_n)
			&\propto (1-\pi_n^{\mathrm{in}})\mathcal{CN}(t_n; \vec{t}_n, \vec{\nu}^t_n)\delta(t_n)\\
			&+ \pi_n^{\mathrm{in}}\mathcal{CN}(t_n; \cev{x}_n, \cev{\nu}^x_n)\mathcal{CN}(t_n; \vec{t}_n, \vec{\nu}^t_n).
		\end{aligned}
	\end{equation}
	Thus, the approximate posterior probability distribution of $t_n$ is given by (\ref{post_pt}), as shown in the top of next page,
	\begin{figure*}[!t]
		\normalsize
		\setcounter{MYtempeqncnt}{\value{equation}}
		\setcounter{equation}{61}
		\begin{equation}
			\begin{aligned}
				p(t_n|\mathbf{r}) &=\frac{\left((1-\pi_n^{\mathrm{in}})\delta(t_n) + \pi_n^{\mathrm{in}}\mathcal{CN}(t_n; \cev{x}_n, \cev{\nu}^x_n)\right) \mathcal{CN}(t_n; \vec{t}_n, \vec{\nu}^t_n)}{\int \left((1-\pi_n^{\mathrm{in}})\delta(t_n) + \pi_n^{\mathrm{in}}\mathcal{CN}(t_n; \cev{x}_n, \cev{\nu}^x_n)\right) \mathcal{CN}(t_n; \vec{t}_n, \vec{\nu}^t_n) \mathrm{d}t_n}\\
				& = (1-\omega_n)\delta(t_n) + \omega_n \mathcal{CN}(t_n; t_n^{\mathrm{tmp}}, \nu_n^{{\mathrm{tmp}}}),
			\end{aligned}
			\label{post_pt}
		\end{equation}
		\hrulefill
		\vspace*{4pt}
	\end{figure*}
	where the auxiliary variables $S_n$, $w_n$, $\nu_n^{{\mathrm{tmp}}}$, and $t_n^{\mathrm{tmp}}$ are respectively given by
	\begin{align}
		\omega_n = & \frac{\pi_n^{\mathrm{in}} S_n}{(1-\pi_n^{\mathrm{in}}) \mathcal{CN}(0; \vec{t}_n, \vec{\nu}^t_n) + \pi_n^{\mathrm{in}} S_n},\\
		\label{Sn} S_n =& \mathcal{CN}(0; \vec{t}_n-\cev{x}_n, \vec{\nu}^t_n+ \cev{\nu}^x_n),\\
		\nu_n^{{\mathrm{tmp}}} = & \frac{\vec{\nu}^t_n  \cev{\nu}^x_n}{\vec{\nu}^t_n + \cev{\nu}^x_n} ,\\
		t_n^{\mathrm{tmp}} =&  \label{tn_tmp}\nu_n^{{\mathrm{tmp}}}\left(\frac{\vec{t}_n}{\vec{\nu}^t_n} + \frac{\cev{x}_n}{\cev{\nu}^x_n} \right).
	\end{align}
	
	According to (\ref{post_pt}), the posterior mean $\hat{t}_n$ and variance $\hat{\nu}^t_n$ are respectively given by
	\begin{equation}
		\begin{aligned}
			 \hat{t}_n  &= \int t_np(t_n|\mathbf{r}) \, \mathrm{d}t_n = \omega_n t_n^{\mathrm{tmp}},\\
			\hat{\nu}^t_n   &= \int \left|t_n -\hat{t}_n \right|^2 p(t_n|\mathbf{r}) \, \mathrm{d}t_n \\
			&=\omega_n\left((1-\omega_n)\left|\hat{t}_n \right|^2 + \nu_n^{{\mathrm{tmp}}} \right).
		\end{aligned}
		\label{hat_t}
	\end{equation}
	Note that due to the fact that $\mathbf{t}$ is equal to $ \mathbf{S}\mathbf{D}\mathbf{c}$, the posterior mean $\hat{\mathbf{t}}$ can be equivalently seen as the approximate posterior estimation of SnS subchannel $\mathbf{h}$.
	
	\begin{remark}
		{Due to the unique prior probability models and factor graph structures \cite{GAMP3,GAMP4}, the message derivations presented from Sec. IV-B  to Sec. IV-D have structural differences from existing works. Unlike the single-layer Gaussian prior in \cite{GAMP4}, we utilize a two-layer hierarchical prior to capture block-sparsity in angular-domain channels. Additionally, our method incorporates a visibility indicator random variable with a Markov chain-based prior, creating inter-layer dependencies that complicate the message update process compared to \cite{GAMP3}. These complexities lead to a factor graph with loops and higher-order dependencies, requiring fundamentally different and tailored message update equations compared to those in \cite{GAMP3,GAMP4}.}
	\end{remark}
	\vspace{-2em}
	\subsection{Messages Approximation in Factor Nodes $f_{z_m}$ and $f_{x_n}$}
	From the factor graphs shown in Fig. \ref{GF}, it is clear that there are $MN$ edges between $f_{z_m}$
	and $t_n$ for all $m$ and $n$. Similarly, there are $NQ$ edges between $f_{x_n}$ and $c_q$ for all $n$ and $q$. Therefore, $2MN+2NQ$ messages have to be updated for forward and backward message passing in each iteration. To further reduce the computational complexity of message passing, we can approximate the means and variances of Gaussian messages by omitting some small terms.
	\subsubsection{Messages approximation for $m_{t_n \rightarrow f_{z_m}}(t_n)$ and $m_{c_q \rightarrow f_{x_n}}(c_q)$}
	According to (\ref{vm_t2ft}), it can be seen that $\vec{\nu}^t_n \ll \vec{\nu}_{m,n}^t$ when $N$ is large enough. Meanwhile, according to the Bayesian theory, we have $\hat{\nu}^t_n < \vec{\nu}^t_n$. Thus, the variance $\cev{\nu}^t_{n,m} = 1/\left(1/\hat{\nu}^t_n- 1/\vec{\nu}^t_{m,n} \right) \approx \hat{\nu}^t_n$. Utilizing (\ref{tmn_r}) and $\cev{\nu}^t_{n,m} \approx \hat{\nu}^t_n$, the mean $\cev{t}_{n,m}$ can be approximated as 
	\begin{equation}
		\begin{aligned}
			\cev{t}_{n,m} &\approx \hat{\nu}^t_n \left(\frac{\hat{t}_n}{\hat{\nu}_n^t} - \frac{a_{m,n}^* \left({r_m - \cev{z}_m} + {a_{m,n}}\cev{t}_{n,m}\right)}{\hat{\beta}^{-1} + \cev{\nu}^z_m- \left|a_{m,n}\right|^2\cev{\nu}^t_{n,m}} \right)\\ &\overset{(a)}{\approx} \hat{t}_n - a_{m,n}^* \hat{\nu}^t_n\frac{({r_m - \cev{z}_m})}{\hat{\beta}^{-1} + \cev{\nu}^z_m},
		\end{aligned}
		\label{tnm}
	\end{equation}
	where $(a)$ is obtained utilizing the fact that $\cev{z}_m \gg {a_{m,n}}\cev{t}_{n,m} $ and $\cev{\nu}^z_m \gg \left|a_{m,n}\right|^2\cev{\nu}^t_{n,m}$ from (\ref{z_mnl}). Similarly, the variance $\cev{\nu}^c_{q,n}$ and mean $\cev{c}_{q,n}$ provided in (\ref{vc_qn}) can be approximated as 
	\begin{align}
		\cev{\nu}^c_{q,n}  \approx \hat{\nu}^c_q,\quad
		\cev{c}_{q,n}     \approx \hat{c}_q - d_{n,q}^* \hat{\nu}^c_q \frac{\vec{x}_n - \cev{x}_n}{\vec{\nu}^x_n + \cev{\nu}^x_n}.
	\end{align}
	\subsubsection{Messages approximation for $m_{f_{z_m} \rightarrow z_m} (z_m)$ and $m_{f_{x_n} \rightarrow x_n}(x_n)$} With the approximated $\cev{\nu}^t_{n,m}$ and $\cev{t}_{n,m}$, the variance $\cev{\nu}^z_m$ and mean $\cev{z}_m$ in (\ref{z_mnl}) can be simplified as 
	\begin{equation}
		\begin{aligned}
			\label{vz_ml}\cev{\nu}^z_m &\approx \sum_{n=1}^{N_{\mathrm{R}}}\left|a_{m,n}\right|^2 \hat{\nu}^t_n,\\
			\cev{z}_m &\approx \sum_{n=1}^{N_{\mathrm{R}}}a_{m,n} \hat{t}_n - \cev{\nu}^z_m\frac{{r_m - \cev{z}_m}}{\hat{\beta}^{-1} + \cev{\nu}^z_m}.	 
		\end{aligned}
	\end{equation}
	Similarly, the variance $\cev{\nu}^x_n$ and mean $\cev{x}_n$ can be also approximated as 
	\begin{equation}
		\begin{aligned}
			\label{vx_nl} \cev{\nu}^x_n  &\approx \sum_{q=1}^Q\left|d_{n,q}\right|^2 \hat{\nu}^c_q,\\
			\cev{x}_n &\approx \sum_{q=1}^Qd_{n,q}\hat{c}_q - \cev{\nu}^x_n \frac{\vec{x}_n - \cev{x}_n}{\vec{\nu}^x_n + \cev{\nu}^x_n}.
		\end{aligned}
	\end{equation}
	\subsubsection{Messages approximation for $m_{t_n \rightarrow f_{t_n}}(t_n)$ in (\ref{vm_t2ft}) and $m_{c_q \rightarrow f_{c_q}}(c_q)$ in (\ref{vc_qr})}
	\begin{equation}
		\begin{aligned}
			\label{t_nr} \vec{\nu}^t_n &\overset{(a)}{\approx} \left(\sum_{m=1}^M \frac{\left|a_{m,n}\right|^2}{\hat{\beta}^{-1} + \cev{\nu}^z_m}\right)^{-1},\\
			\vec{t}_n &
			\overset{(b)}{\approx} \hat{t}_n + \vec{\nu}^t_n\sum_{m=1}^Ma_{m,n}^*\frac{{r_m - \cev{z}_m}}{\hat{\beta}^{-1} + \cev{\nu}^z_m},
		\end{aligned}
	\end{equation}
	where $(a)$ are obtained with $\cev{\nu}^z_m \gg \left|a_{m,n}\right|^2\cev{\nu}^t_{n,m}$; $(b)$ is obtained with ${\left|a_{m,n}\right|^2}/({\hat{\beta}^{-1} + \cev{\nu}^z_m}) \ll 1/\hat{\nu}^t_n$, since ${\left|a_{m,n}\right|^2}/({\hat{\beta}^{-1} + \cev{\nu}^z_m}) \ll 1/\vec{\nu}^t_n$ for large $M$ from (\ref{t_nr}) and $\vec{\nu}^t_n \ge \hat{\nu}^t_n$.
	Similarly, the variance $\vec{\nu}^c_q$ and mean $\vec{c}_q$ in (\ref{vc_qr}) can be approximated as 
	\begin{equation}
		\begin{aligned}
			\label{vc_qr2} \vec{\nu}^c_q &\approx \left(\sum_{n=1}^{N_{\mathrm{R}}} \frac{\left|d_{n,q}\right|^2}{\vec{\nu}^x_n + \cev{\nu}^x_n}\right)^{-1},\\
			 \vec{c}_q &\approx \hat{c}_q + \vec{\nu}^c_q\sum_{n=1}^{N_{\mathrm{R}}}d_{n,q}^*\frac{\vec{x}_n - \cev{x}_n}{\vec{\nu}^x_n + \cev{\nu}^x_n}. 
		\end{aligned}
	\end{equation}
	\vspace{-2em}
	\subsection{Overall Algorithm and Complexity Analysis}
	The proposed TL-GAMP algorithm can be organized in a more succinct form, which is summarized in Algorithm~\ref{TL-GAMP} and it can be terminated when it reached a maximum number of iteration or the difference between the estimates of two consecutive iterations is less than a
	threshold. 
	Once the subchannels $\hat{\mathbf{t}}_l$ for $l=1,2,\cdots, L$ are obtained, the final channel is estimated as $\hat{\mathbf{H}} = \sum_{l=1}^L\hat{\mathbf{t}}_l\mathbf{a}_{\mathrm{T}}^{\mathrm{H}}(\hat{\psi}_l)$. 
	\begin{remark}
		{Compared to the existing SnS channel estimation algorithms in \cite{Bayesian1, Turbo-OMP}, the proposed TL-GAMP algorithm simultaneously achieves VR detection in the antenna domain and channel estimation in the angular domain through a three-layer Bayesian inference architecture. On one hand, the incorporation of angular-domain sparsity enhances channel estimation by providing additional channel characteristics. On the other hand, the concurrent VR detection and channel estimation prevent error propagation, a limitation of the two-stage Turbo-OMP algorithm \cite{Turbo-OMP}, resulting in improved estimation performance with the TL-GAMP algorithm.}
	\end{remark}
	
	\begin{algorithm}
		\renewcommand{\algorithmicrequire}{\textbf{Input:}}
		\renewcommand{\algorithmicensure}{\textbf{Output:}}
		\caption{TL-GAMP algorithm for subchannel estimation}
		\begin{algorithmic}[1]
			\Require received vector $\mathbf{r}$, measurement matrix $\mathbf{A}$ and codebook $\mathbf{D}$; 
			\Statex \textbf{Initialize:} $\left\{ \vec{\nu}^c_q,  \vec{c}_q, \forall q \right\}$, $\left\{\vec{t}_n, \vec{\nu}_n^t, \hat{t}_n, \hat{\nu}^t_n, \forall n\right\}$, $\left\{\cev{z}_m, \cev{\nu}^z_m, \forall m \right\}$, $\xi$, $\eta$, $p_{01}$, $\psi_0^f$, $\psi^b_{N_{\mathrm{R}}}$ and  $\hat{\beta}$. 
			\While{the stopping criterion is not met}
			\Statex /*\textbf{The first layer}*/
			\State Update the posterior estimate of $\gamma_q$ according to (\ref{gamma});
			\State Update the posterior estimate of $c_q$ according to (\ref{v_cq});
			
			\State Update the output message $m_{f_{x_n} \rightarrow x_n}(x_n)$ of the first layer according to (\ref{vx_nl});
			\State Update the message from $c_q$ to $f_{c_q}$ according to (\ref{vc_qr2});
			\Statex /*\textbf{The second layer}*/
			\State Update the forward and backward messages of Markov chain according to (\ref{psi_f}) and (\ref{psi_b});
			\State Update the output messages of Markov chain according to (\ref{pi_in});
			
			\State Update the belief of $s_n$ according to (\ref{bsn});
			
			\State Update $\psi^f_{0}$ of Markov chain according to (\ref{psi_0});
			\State Update the output message of the second layer to $x_n$ given by $\vec{x}_n = \vec{t}_n$, $\vec{\nu}^x_n = \vec{\nu}^t_n$;
			\Statex /*\textbf{The third layer}*/
			\State Update message from $z_m$ to $f_{z_m}$ according to (\ref{vz_ml});
			\State Update posterior estimate of $\beta$ according to (\ref{post_beta});
			\State Update posterior estimate of $z_m$ according to (\ref{vz_m});
			\State Update message from $t_n$ to $f_{t_n}$ according to (\ref{t_nr});
			\State Update posterior estimation $t_n$ according to (\ref{hat_t}).
			\EndWhile
			\Ensure $\hat{t}_n$ for all $n$.
		\end{algorithmic}
		\label{TL-GAMP}
	\end{algorithm}
	
	In the following, we provide the computational complexity analysis for the TL-GAMP algorithm. The TL-GAMP algorithm requires pre-processing, i.e., performing an economic SVD for $\mathbf{P}$ and unitary transformation, and the complexity is $\mathcal{O}(M^2N_{\mathrm{R}})$. It is noted that the pre-processing can be carried out offline, thus, the computational complexity can be overlooked\footnote{{Since $\mathbf{P}$ depends solely on the matrices $\mathbf{B}$ and $\mathbf{W}$ and is independent of the estimation results from Phase I. Thus, the SVD for matrix $\mathbf{P}$ only needs to be computed when the noise covariance matrix $\mathbf{R} = \mathbf{B}\mathbf{B}^{\mathrm{H}}$ changes. Since the second-order statistical characteristic $\mathbf{R}$ of the noise vector varies more slowly compared to the instantaneous channel information $\mathbf{H}$, the SVD for matrix $\mathbf{P}$ does not need to be performed within every coherence block. Instead, it is required only when $\mathbf{R}$ changes. By averaging the computational complexity of the SVD over multiple coherence blocks, its impact becomes negligible.}}.
	Examining the TL-GAMP algorithm, it is evident that there is no matrix inversion involved, and the most computationally intensive parts only involve matrix-vector products. Specifically, in the first layer, the complexity of the proposed algorithm is dominated by the computation of $\vec{c}_q$ and $\cev{x}_n$ for all $q$ and $n$, which requires $\mathcal{O}(N_{\mathrm{R}} Q)$. In the second layer, the complexity is dominated by the computation of $\pi_n^{\mathrm{out}}$, $\pi_n^{\mathrm{in}}$, $\psi_n^{f}$, $\psi_n^{b}$, and $b(s_n)$, which requires $\mathcal{O}(N_{\mathrm{R}})$. In the third layer, the complexity is dominated by the computation of $\vec{t}_n$ and $\cev{z}_m$, given by $\mathcal{O}(M N_{\mathrm{R}})$. Therefore, the complexity of TL-GAMP per iteration is $\mathcal{O}(N_{\mathrm{R}}Q+MN_{\mathrm{R}})$, which linearly increases with $M$, $N_{\mathrm{R}}$, and $Q$.
	For comparison, we also provide the overall complexity of Turbo-BOMP \cite{Turbo-OMP}, which is $\mathcal{O}(I_1 MN_{\mathrm{R}} +  I_2N_{\mathrm{R}}Q)$, where $I_1$ and $I_2$ denote the number of iterations in stages I and II, respectively. It can be seen that the two algorithms have comparable computational complexity. However, it is important to note that the novel TL-GAMP algorithm exhibits significant performance superiority compared to the Turbo-BOMP algorithm, as will be verified in Section~\ref{section5}.
	\vspace{-1em}
	\section{Simulation Results}
	\label{section5}
	\begin{table}
		\centering
		\caption{Simulation Parameters}
		\setlength{\tabcolsep}{4mm}{
			\begin{tabular}{c c}
				\toprule
				\makecell[c]{Notations}& Parameters\\ 
				\midrule
				\makecell[l]{Number of BS antennas $N_{\mathrm{R}}$} &256 \\
				\makecell[l]{Number of User antennas $N_{\mathrm{T}}$} &16\\
				\makecell[l]{Number of RF chains $N_{\mathrm{RF}}$} &8\\
				\makecell[l]{Carrier frequency $f_c$}    &30GHz\\
				\makecell[l]{Number of channel paths $L$} &4\\
				\makecell[l]{Angle of arrival $\vartheta_l$} &$\mathcal{U}(-\pi/2, \pi/2)$\\
				\makecell[l]{Angle of departure $\psi_l$} &$\mathcal{U}(-\pi/2, \pi/2)$\\
				\makecell[l]{Distance between BS and UE or scatterers $r_l$} &[2, 10]m\\
				\makecell[l]{Number of channel realizations} &2000\\
				\bottomrule
			\end{tabular}
		}
		\label{Parameters}
	\end{table}
	
	In this section, we evaluate the performance of the proposed channel estimation scheme under various simulations. The system parameters are shown in Table \ref{Parameters}. {Notably, in the all simulation results, the different paths correspond to different VRs, as shown in Fig. \ref{System Model}.}
	To effectively evaluate estimation performance, we consider the normalized mean square error (NMSE) as the performance metrics, which is defined as $\mathrm{NMSE} \triangleq {\lVert \hat{\mathbf{H}} - \mathbf{H} \rVert^2_{\mathrm{F}}}/{\lVert \mathbf{H} \lVert^2_{\mathrm{F}}}$,
	where $\mathbf{H}$ and $\hat{\mathbf{H}}$ are the true channel and estimated channel, respectively. 
	{The SNR is defined at the receive, and is given by $10\log_{10}\left(\lVert\mathbf{W}^{\mathrm{H}}\mathbf{H}\mathbf{f}_p \rVert^2_{\mathrm{F}}/ \lVert\overline{\mathbf{n}}_p \rVert^2_{\mathrm{F}}\right)$ in dB.}
	Additionally, we compare the proposed TL-GAMP algorithm with the following benchmarks:
	\begin{itemize}
		\item \textbf{LS}: Least squares estimator based on the formulation (\ref{pilot3}). More specifically, the SnS subchannel is estimated as $\hat{\mathbf{h}}_{l,\mathrm{LS}}=(\mathbf{W}\mathbf{W}^{\mathrm{H}})^{-1}\mathbf{W}^{\mathrm{H}}\mathbf{y}_l$ for all $l$.
		\item \textbf{PD-OMP}: On-grid polar-domain simultaneous orthogonal matching pursuit algorithm for NF channels proposed in \cite{PolarCS} without considering the SnS properties.
		
		\item \textbf{AD-GAMP}: Antenna-domain generalized approximate message passing algorithm without utilizing the angular-domain statistical characteristics, which is similar to the methods proposed in \cite{Bayesian1,Bayesian3}.
		\item \textbf{Turbo-BOMP}. Two-stage VR detection and channel estimation scheme proposed in our previous work \cite{Turbo-OMP}, where the angular-domain information can not be fully utilized in VR detection stage.
		\item {\textbf{P-BiG-AMP:} Parametric bilinear generalized approximated message passing \cite{Para_Bi_Est}, where $\mathbf{s}_l$  and $\mathbf{c}_l$ were drawn from i.i.d. Bernoulli-$\mathcal{CN}(0, \nu^{s}_l)$ and Bernoulli-$\mathcal{CN}(0, \nu^{c}_l)$ with sparsity rates $\epsilon^s_l$ and $\epsilon^c_l$, respectively.}
		\item \textbf{TL-GAMP with perfect AoDs:} TL-GAMP algorithm with perfect AoDs knowledge to evaluate the impact of AoDs estimation accuracy on subsequent channel estimation in Phase II.
		\item \textbf{TL-GAMP with perfect VR:} TL-GAMP algorithm with perfect VR knowledge as a performance lower bound.
	\end{itemize}

	\begin{figure*}
		\centering
		\begin{minipage}{0.45\linewidth}
			\centering
			\includegraphics[width=0.93\linewidth]{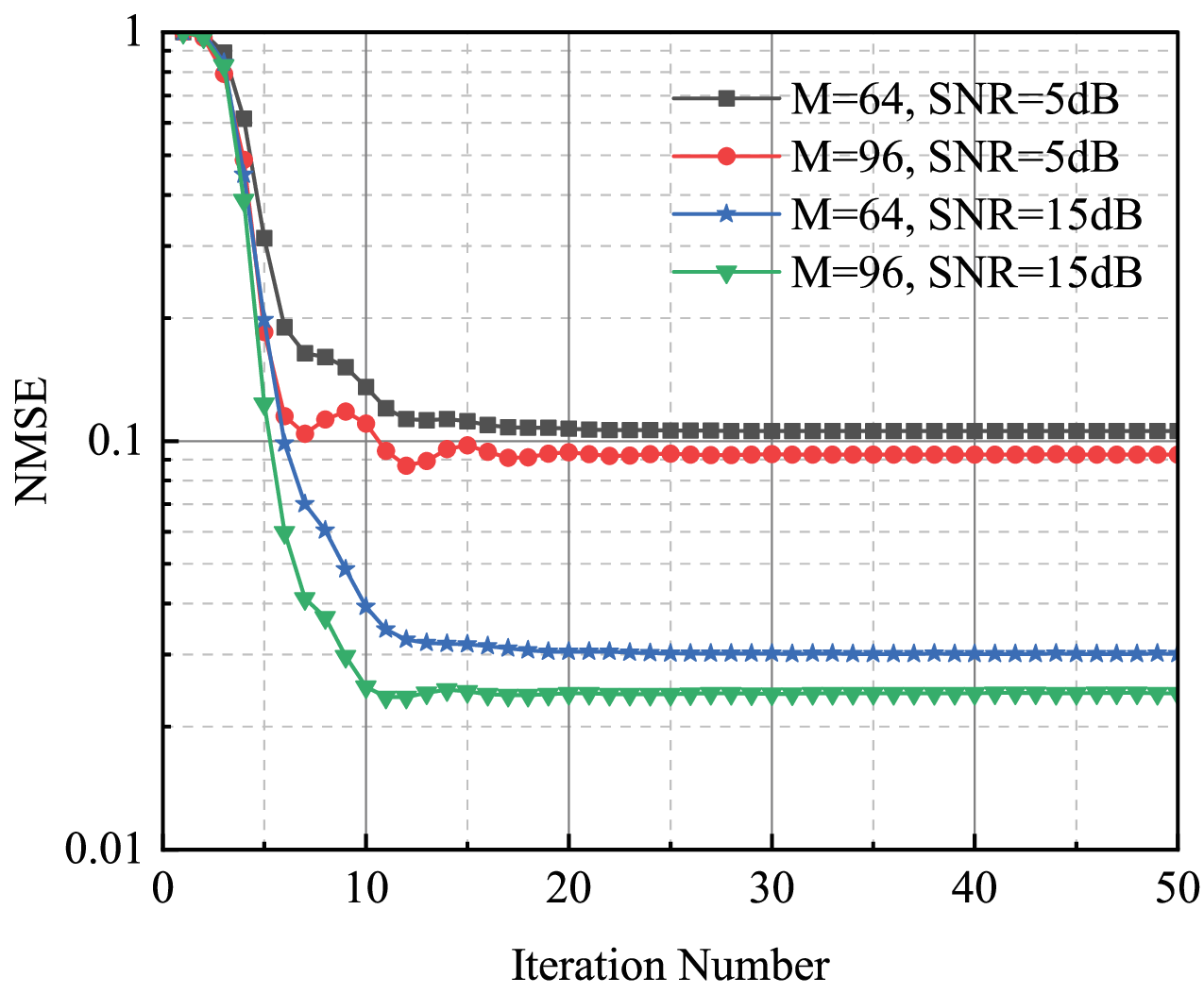}
			\caption{Convergence performance.}
			\label{convergence}
		\end{minipage}
		\qquad
		\begin{minipage}{0.45\linewidth}
			\centering
			\includegraphics[width=0.9\linewidth]{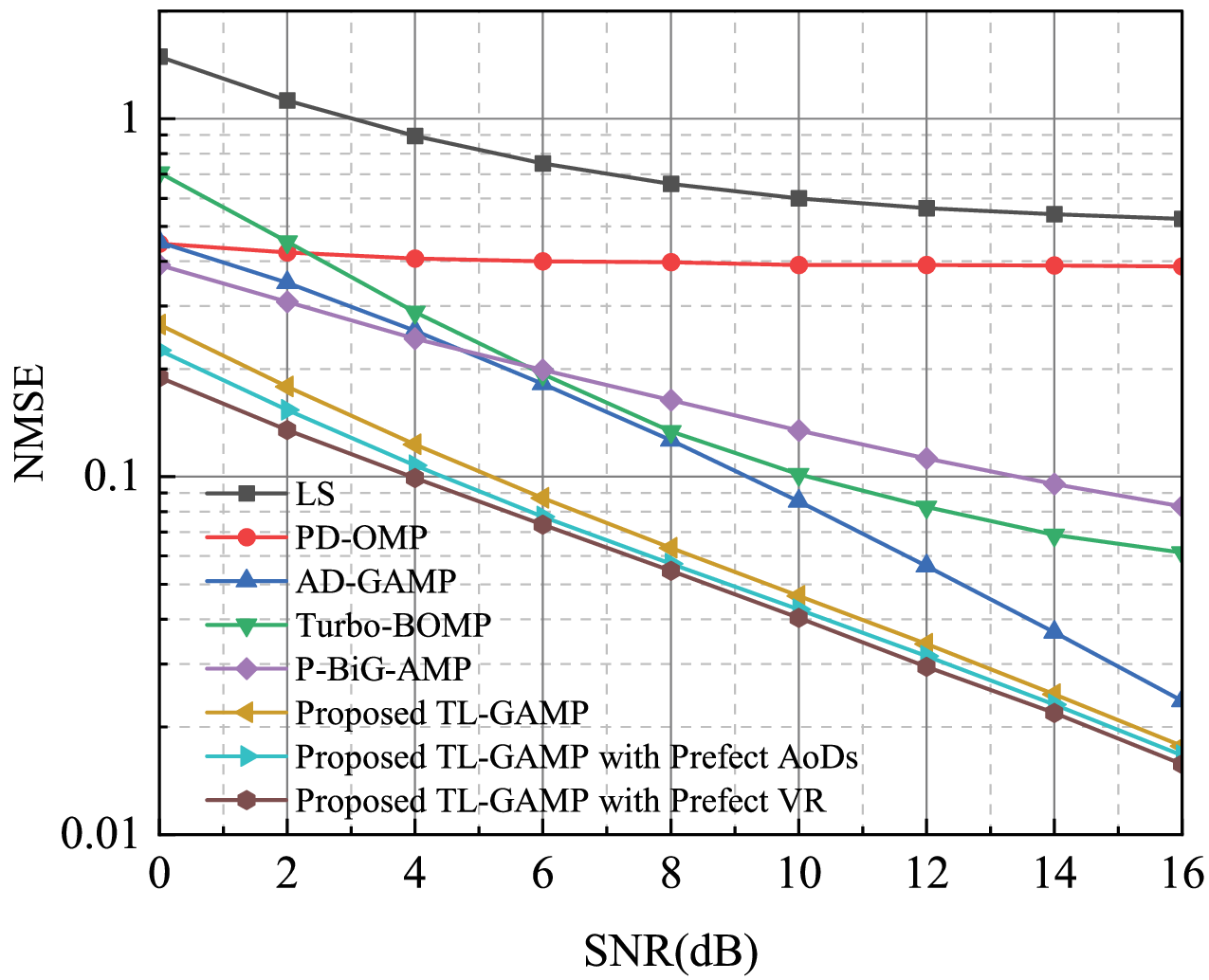}
			\caption{NMSE versus SNR with VR size of 0.25.}
			\label{rho1}
		\end{minipage}
	\end{figure*}
	
	\begin{figure*}
		\centering
		\begin{minipage}{0.45\linewidth}
			\centering
			\includegraphics[width=0.90\linewidth]{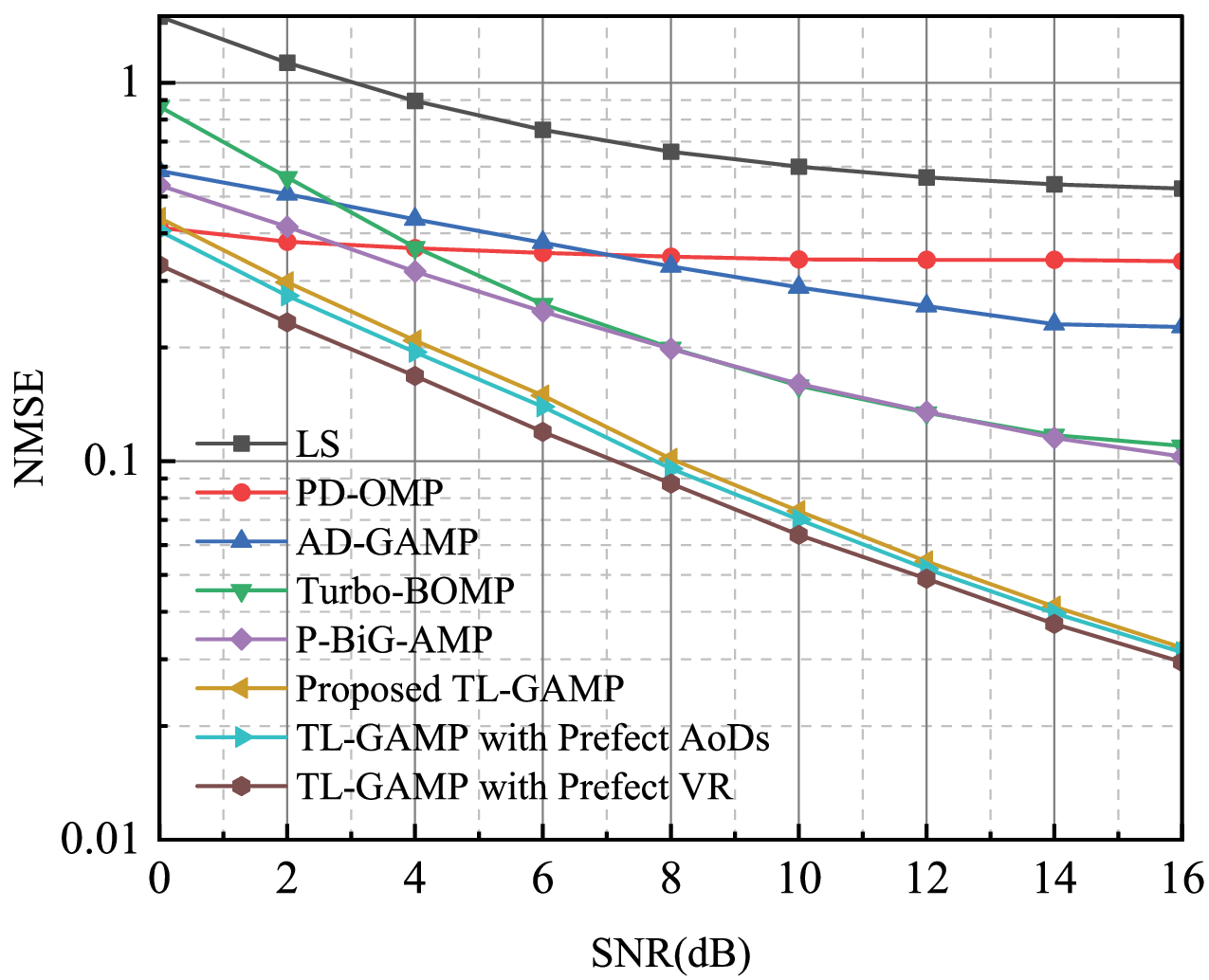}
			\caption{NMSE versus SNR with VR size from 0.2 to 1.}
			\label{rho2}
		\end{minipage}
		\qquad
		\begin{minipage}{0.45\linewidth}
			\centering
			\includegraphics[width=0.9\linewidth]{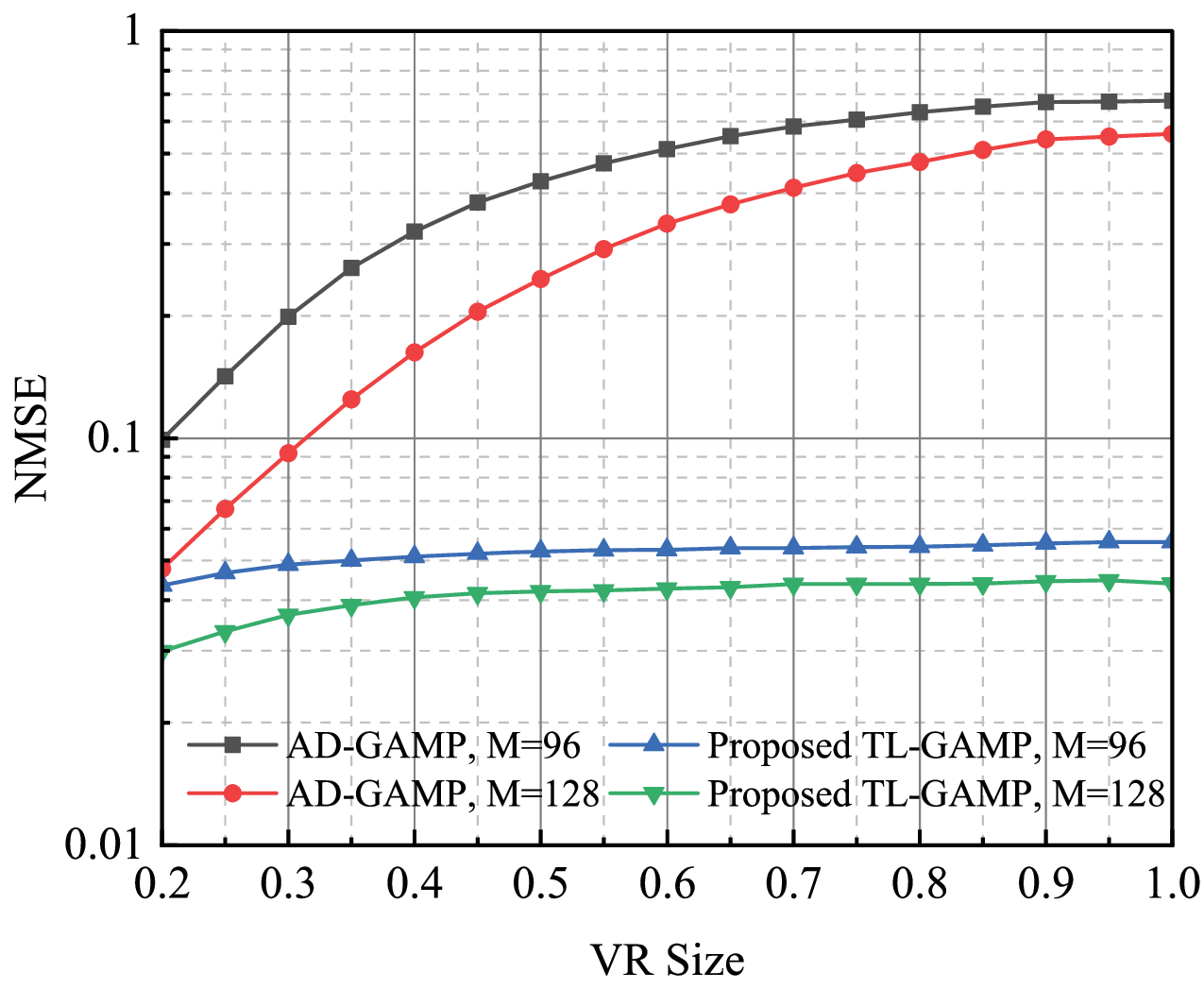}
			\caption{NMSE versus VR size.}
			\label{psi}
		\end{minipage}
	\end{figure*}
	{Fig. \ref{convergence} illustrates the average convergence behavior of the TL-GAMP algorithm, plotting the NMSE against the iteration number under varying pilot lengths and SNR conditions. The results demonstrate a consistent monotonic decrease in NMSE across all iterations for the different simulation settings, confirming the stable convergence of the proposed TL-GAMP algorithm.}
	Considering both performance and computational cost, the maximum iteration can be safely set as 20 for subsequent simulations.
	
	{Fig. \ref{rho1} investigates the NMSE performance versus SNR with fixed VR size of $\phi_l = 0.25$ and $M=128$. The results highlight the importance of extracting SnS properties in ensuring accurate estimation performance. Specifically, ignoring SnS characteristics, as seen in methods like LS and PD-OMP, leads to significant performance degradation.
	Meanwhile, the proposed TL-GAMP algorithm demonstrates significant superiority over existing methods that do account for SnS properties, such as AD-GAMP, Turbo-BOMP, and P-BiG-AMP algorithm. Unlike the AD-GAMP approach, which relies solely on spatial-domain channel characteristics, TL-GAMP algorithm simultaneously leverages correlations in the spatial domain and sparsity in the angular domain, resulting in enhanced estimation accuracy. Compared to the Turbo-BOMP approach, which exists error propagation due to its two-stage process, the TL-GAMP algorithm improves estimation performance by performing concurrent VR detection and channel estimation. Furthermore, the TL-GAMP algorithm outperforms the P-BiG-AMP method by incorporating structured prior probability models that are specifically tailored to capture the characteristics of SnS-NF channels.} 
	Furthermore, the NMSE performance of the proposed TL-GAMP algorithm closely approaches the lower bounds set by perfect AoD and VR knowledge. This convergence indicates that the AoD estimation and VR detection effectively capture the channel characteristics, achieving near-optimal performance and validating the robustness and accuracy of the overall estimation scheme.
	
	{More generally, Fig. \ref{rho2} presents the NMSE performance as a function of SNR for varying VR sizes ranging from 0.2 to 1, with $M = 128$. A notable observation is that, unlike the case when $\phi_l = 0.25$, the AD-GAMP method experiences a significant performance drop as the VR size increases. This divergence in performance is due to the AD-GAMP algorithm's reliance on spatial-domain sparsity; as the VR expands, the sparsity in the spatial domain diminishes, leading to corresponding performance degradation. In contrast, the TL-GAMP algorithm demonstrates remarkable robustness to changes in VR size.} 
	To further substantiate this observation, Fig. \ref{psi} plots the NMSE performance against VR size $\phi_l$ with $\mathrm{SNR} = 10$ dB. 
	It is evident that the proposed TL-GAMP algorithm only exhibits slight fluctuations in performance as the VR size changes, indicating that the joint utilization of statistical characteristics in both the antenna and angular domains significantly enhances the algorithm's robustness to VR size. Consequently, the proposed TL-GAMP algorithm is well-suited for both NF-SnS and NF-SS scenarios.
	
	{Fig. \ref{pilot} compares the performance of the proposed TL-GAMP algorithm with various benchmarks across different pilot lengths $M$ under $\mathrm{SNR} = 10$ dB and $\phi_l \in [0.1, 1]$ for all $l$. The pilot length $M$ varies from $48$ to $192$. It is observed that the performance of all algorithms improves as the pilot length $M$ increases. Notably, as the pilot length grows, the rate of performance improvement gradually approaches saturation. Therefore, a moderate value of $M$ can be selected to balance performance and pilot overhead.
	Furthermore, the proposed TL-GAMP algorithm significantly outperforms the existing baselines across almost the entire considered range. Thus, the TL-GAMP algorithm offers a low-overhead SnS XL-MIMO channel estimation scheme.}
	Additionally, it can be observed that the proposed TL-GAMP algorithm consistently maintains a small gap relative to the lower bounds obtained by perfect AoD and VR knowledge. Notably, as the pilot length increases, TL-GAMP approaches the optimal performance with perfect VR information. This trend indicates that an extended pilot length significantly enhances VR estimation accuracy, drawing it closer to the ideal scenario.
	\begin{figure}
		\centering
		\includegraphics[width=0.4\textwidth]{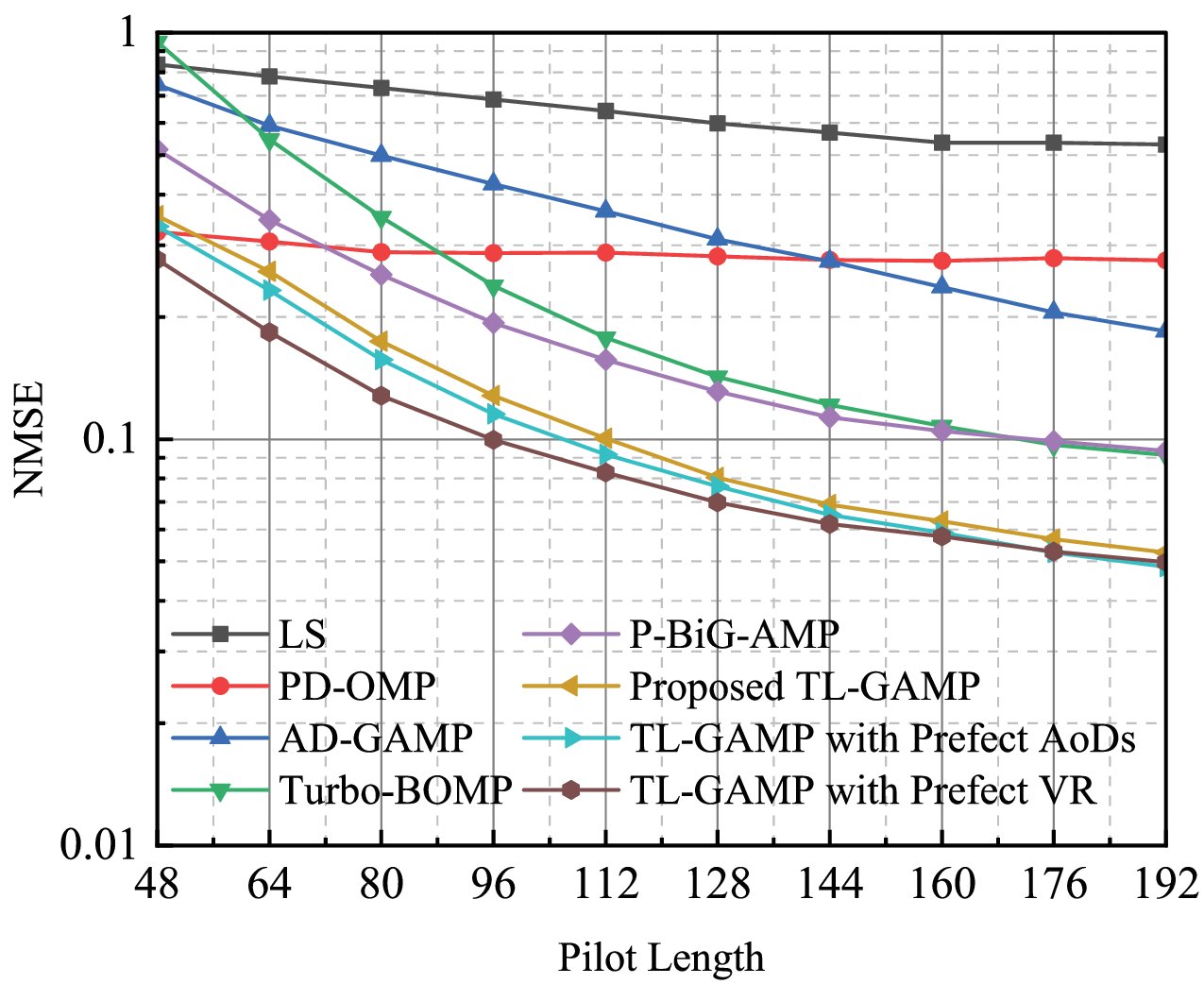}
		\caption{NMSE versus pilot length.}
		\label{pilot}
	\end{figure} 
	\begin{figure}
		\centering
		\includegraphics[width=0.4\textwidth]{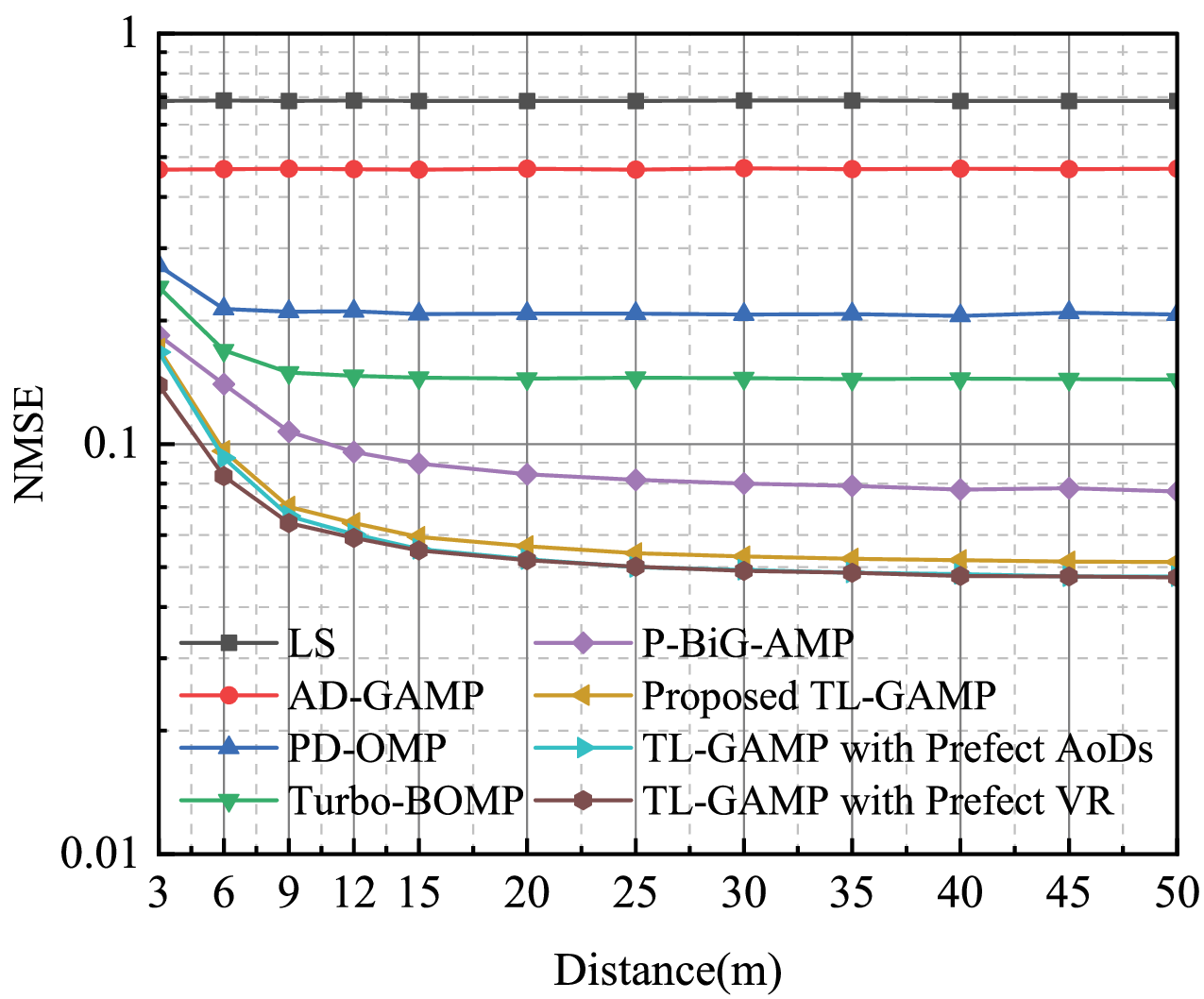}
		\caption{NMSE versus distance between BS and scatterers.}
		\label{distance}
	\end{figure} 
	
	Fig. \ref{distance} illustrates the NMSE performance as a function of the distance between the BS and scatterers, with fixed values of $M = 96$ and $\mathrm{SNR} = 10$ dB. The results reveal a slight performance degradation as the distance decreases, particularly within the $3$ to $15$ meters range. This degradation is attributed to an increase in significant spatial frequency components associated with NF channels as the distance decreases, which reduces the sparsity level of $\mathbf{c}_l$, thereby impacting the estimation performance. Notably, the estimation performance gradually stabilizes when the distance exceeds $15$ meters. Overall, the proposed TL-GAMP algorithm demonstrates remarkable robustness across varying distances, underscoring its suitability for both NF and FF scenarios.
	
	\begin{figure*}
		\centering
		\subfigure[]{
			\includegraphics[width=0.3\textwidth]{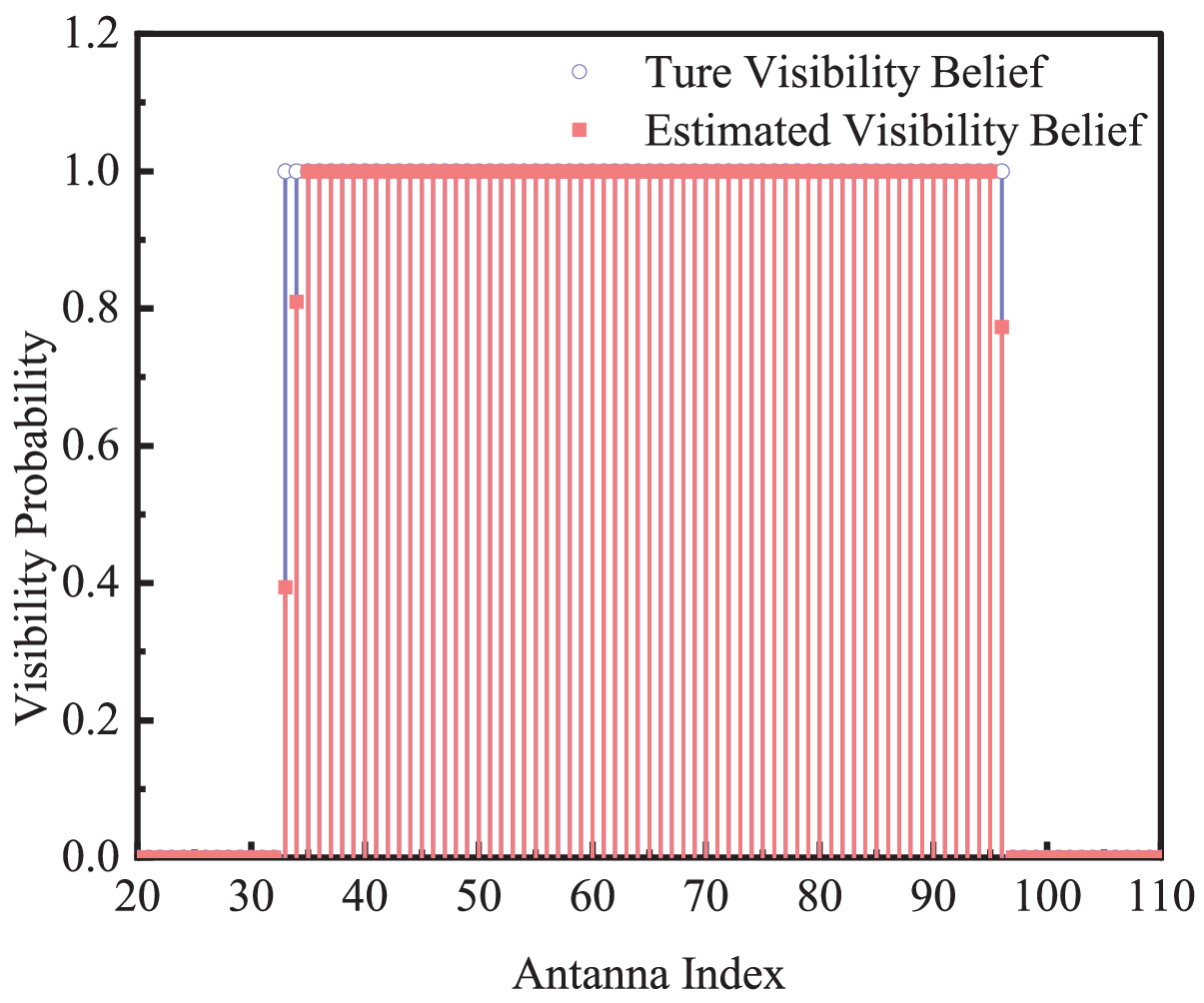}
			\label{vr1}
		}
		\quad
		\subfigure[]{
			\includegraphics[width=0.3\textwidth]{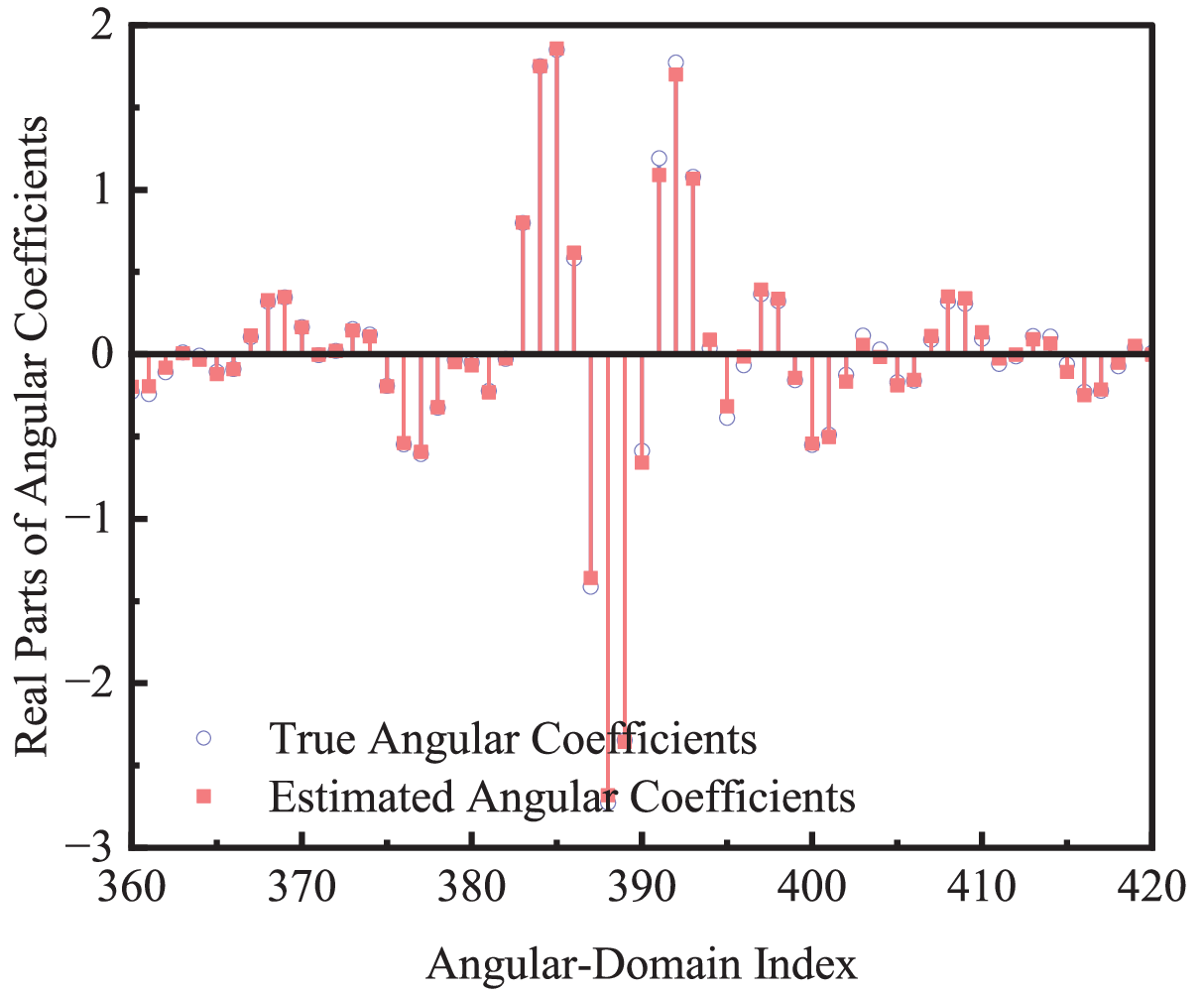}
			\label{vr2}
		}
		\quad
		\subfigure[]{
			\includegraphics[width=0.3\textwidth]{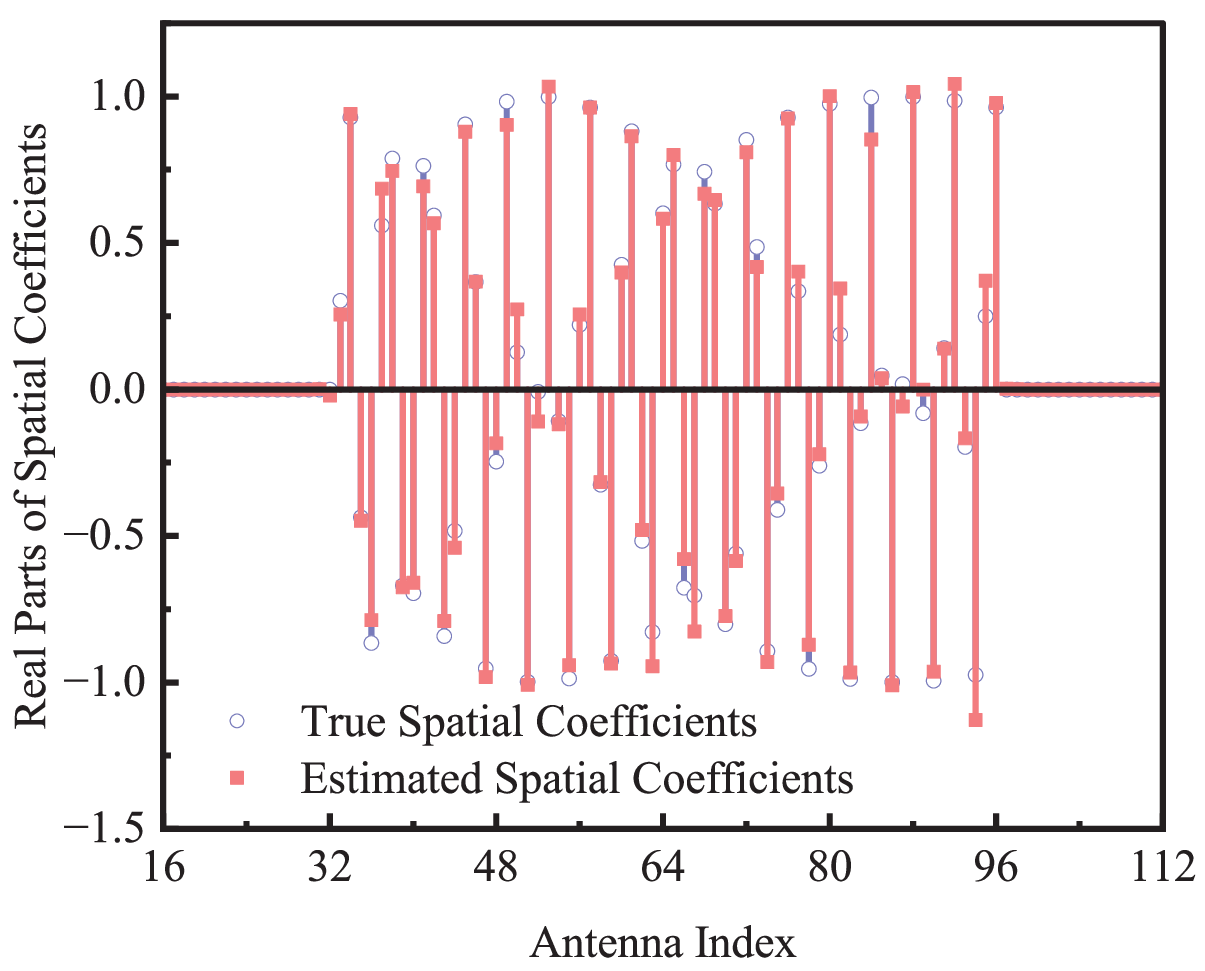}
			\label{vr3}
		}
		\subfigure[]{
			\includegraphics[width=0.29\textwidth]{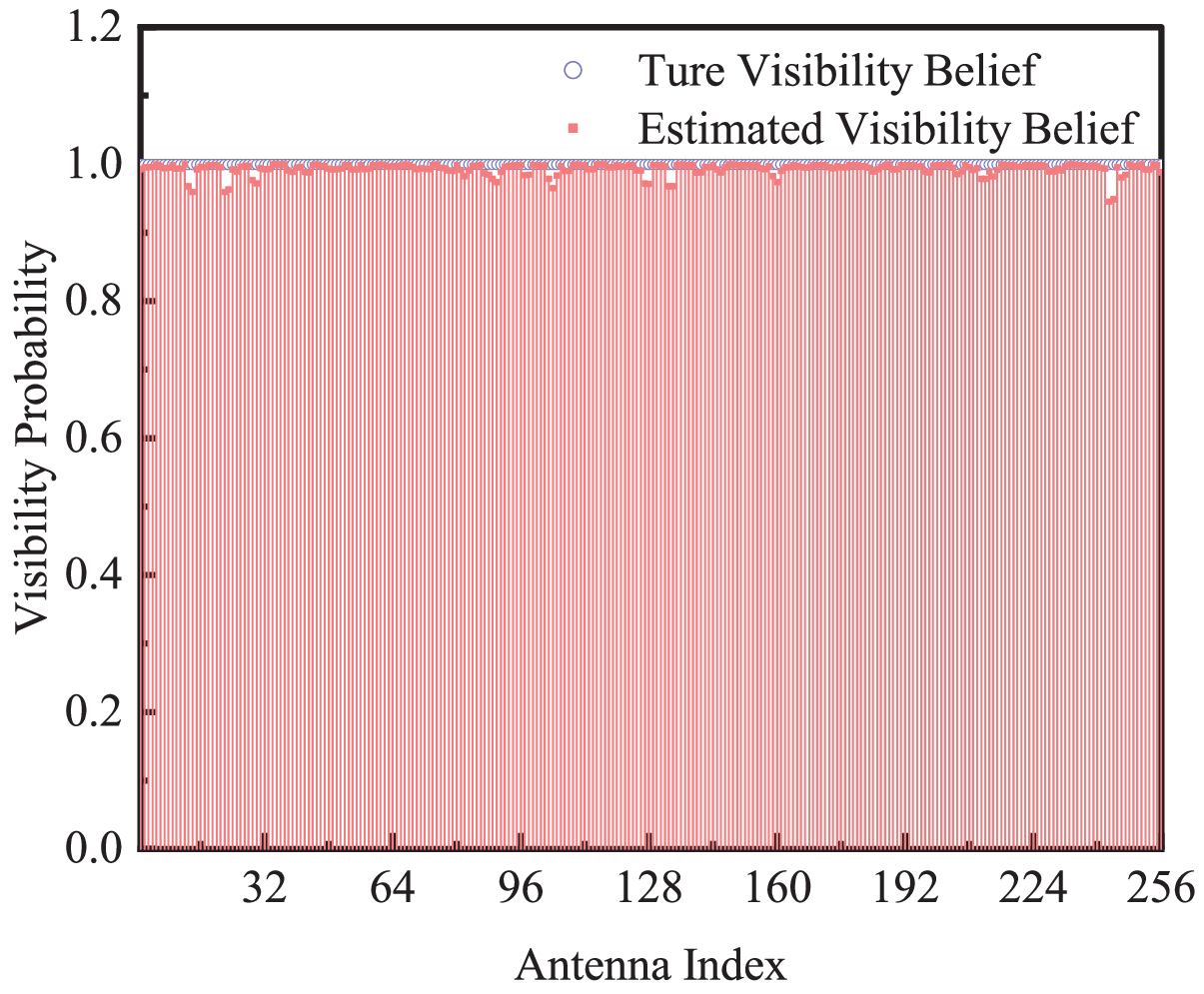}
			\label{c1}
		}
		\quad
		\subfigure[]{
			\includegraphics[width=0.29\textwidth]{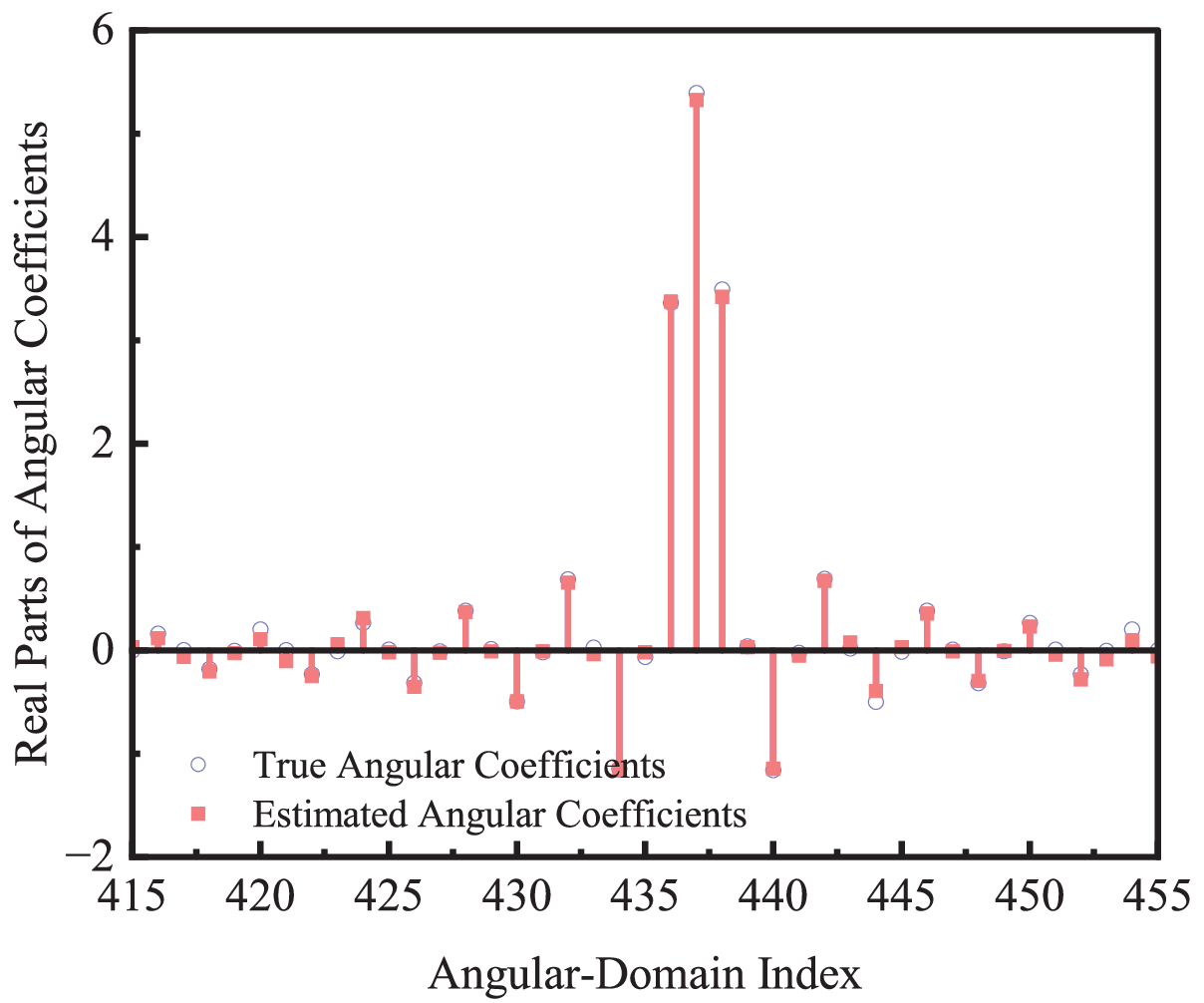}
			\label{c2}
		}
		\quad
		\subfigure[]{
			\includegraphics[width=0.31\textwidth]{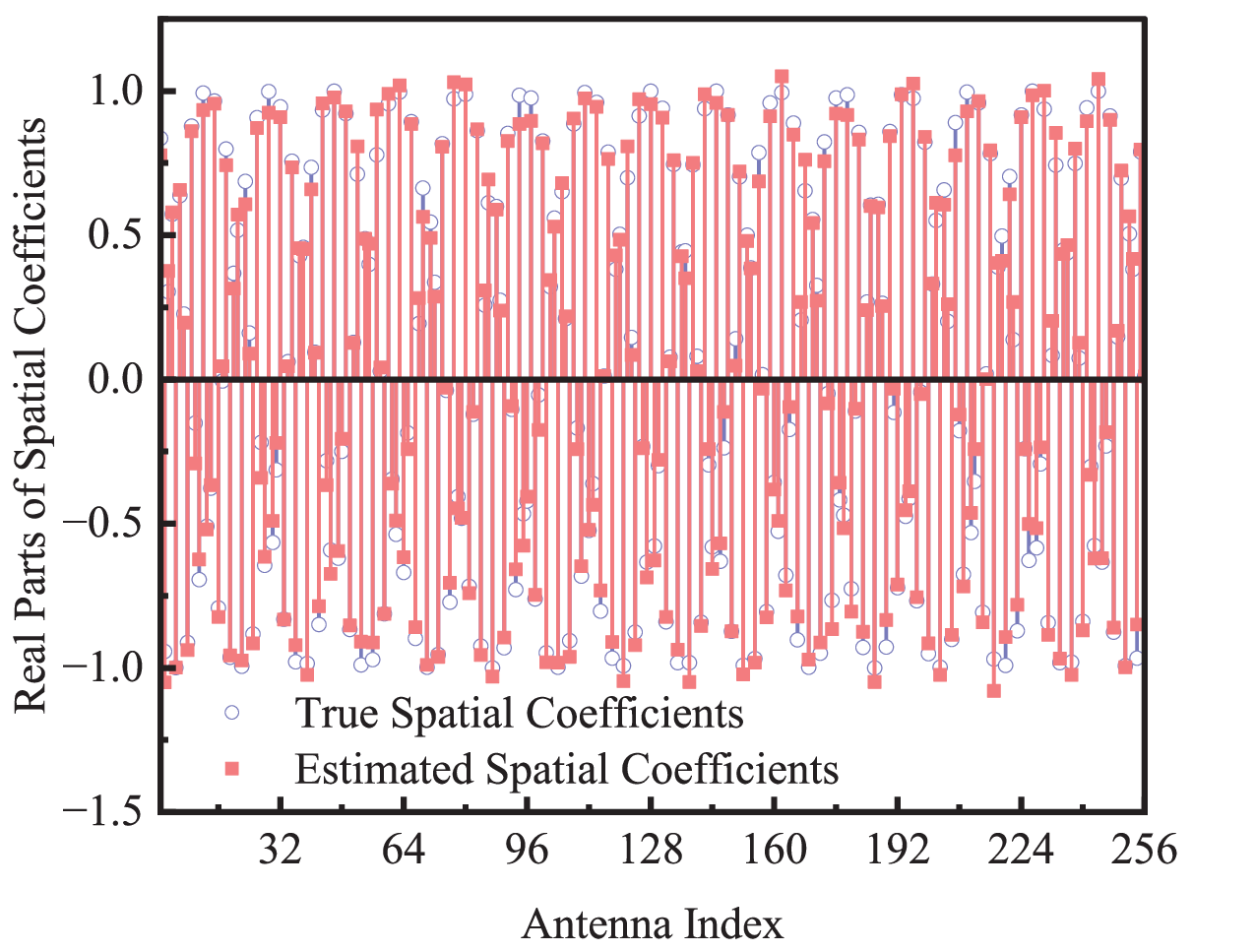}
			\label{c3}
		}
		\caption{Illustration of the estimation results under NF-SnS and FF-SS scenarios. (a) Estimated visibility belief with VR size of 0.25; (b) Estimated angular channel coefficients with VR size of 0.25; (c) Estimated spatial channel coefficients with VR size of 0.25; (d) Estimated visibility belief with VR size of 1; (e) Estimated angular channel coefficients with VR size of 1; (f) Estimated spatial channel coefficients with VR size of 1.}
		\label{VR}
	\end{figure*}
	To demonstrate the estimation performance more intuitively, Fig.~\ref{VR} presents the Monte Carlo simulation results for both NF-SnS and FF-SS scenarios, with $\mathrm{SNR}=15$dB and $M=128$. For the NF-SnS path, the distance between the BS and the scatterer is set to 10m, with the VR size set to 0.25. For the FF-SS path, the distance is 200m, and the VR size is 1. The figure compares the true visibility belief, angular channel coefficients, and spatial channel coefficients of a channel realization with those reconstructed by the TL-GAMP algorithm. It is evident that, regardless of whether the scenario is NF-SnS or FF-SS, the proposed TL-GAMP algorithm consistently demonstrates robust VR detection and accurate estimation of the corresponding channel coefficients.
	\section{Conclusion}
	\label{section6}
	In this paper, we have addressed the channel estimation problem in XL-MIMO systems, considering the spherical wavefront effects and SnS properties. 
	To effectively tackle the joint tasks of channel estimation and VR detection, we propose a two-phase algorithm that strategically decouples the problem into multiple subchannel estimation tasks. 
	For each subchannel estimation, we introduce an efficient TL-GAMP algorithm that leverages the characteristics of SnS subchannels in both the antenna and angular domains. 
	Simulation results indicate that the proposed TL-GAMP algorithm consistently outperforms existing methods, closely approaching the lower bound obtained with perfect AoD and VR knowledge.
	Additionally, the algorithm's robustness in handling NF-SnS, NF-SS, and FF-SS scenarios has been validated.
	\bibliographystyle{IEEEtran}
	\bibliography{ref}
\end{document}